\documentclass[preprint,nofootinbib,aps,superscriptaddress,eqsecnum]{revtex4-2} 
\usepackage[colorlinks,citecolor=red]{hyperref}
\usepackage[usenames]{color}
\usepackage{xcolor}
\usepackage{float}
\pdfoutput=1
\usepackage{tabularx}
\usepackage{graphicx}
\usepackage{amsmath}
\usepackage{amsfonts}
\usepackage{amssymb}
\usepackage[justification=centering]{caption} 
\usepackage{caption}
\usepackage{subcaption}
\usepackage{color}
\allowdisplaybreaks
\def\bea{\begin{eqnarray}}
	\def\eea{\end{eqnarray}}
\def\be{\begin{equation}}
	\def\ee{\end{equation}}

\begin{document}
	\title{Neutrino mass textures and associated phenomenology in modular Left-Right Symmetric Model}
	\author{Ankita Kakoti}
	\email{ankitak@tezu.ernet.in}
	\affiliation{Department of Physics, Tezpur University, Tezpur 784028, India}
	
	\author{Happy Borgohain}
	\email{haps.tezu@gmail.com}
	\affiliation{Department of Physics, Silapathar College, Silapathar, 787059, India}
	
	\author{Mrinal Kumar Das}
	\email{mkdas@tezu.ernet.in}
	\affiliation{Department of Physics, Tezpur University, Tezpur 784028, India}
	\begin{abstract}
		
			Neutrino mass textures play a crucial role in the study of various neutrino mass models and the associated phenomenology. The present work focuses on neutrino phenomenology in the framework of left-right symmetric model (LRSM) augmented by $A_4$ modular symmetry. More specifically we concentrated on  the implementation of  modular  group of level 3 ($\Gamma(3)$). As the use of modular symmetry demands the assignment of different modular weights to the particle content of the model, we consider modular weights  $k_Y$ = 4, 6, 8, 10 in LRSM which gives rise to different neutrino textures (texture zero structures) and  study its consequent neutrino phenomenologies. Observables like resonant leptogenesis (RL), new physics contributions to neutrinoless double beta decay  (NDBD)(momentum dependent ($\lambda$ and $\eta$) and right-handed neutrino contributions of NDBD and lepton flavor violation (LFV) has been deliberated. The study has been carried out for both normal and inverted ordering of neutrino mass and the correlations among the neutrino parameters for the present model has been elaborated for the TeV scale LRSM which can be tested in the experiments.

	\end{abstract}
	\maketitle
	\newpage
	\begin{center}
		\section{\label{lrsm11}\textbf{Introduction}}
	\end{center}
The most celebrated Standard Model (SM) of particle physics fails to explain some of the important phenomena in the neutrino sector, like the generation of neutrino mass, the origin of baryon asymmetry in the universe, origin of dark matter etc. The nature of neutrinos (Dirac or Majorana), the hierarchy (normal or inverted) are also some issues in which SM cannot shed any light upon. This brings into picture the Beyond Standard Model framework, which suggests that the simplest mechanism to generate neutrino masses is the seesaw mechanism, which can be stated as the straightforward way to understand the underlying principle of generation of tiny neutrino mass. The two main components of the mechanism are addition of right-handed (RH) neutrinos to the SM and endowing the RH neutrinos with a Majorana mass which breaks the accidental global (B-L) symmetry of the SM. It is a well known fact that there are two ultraviolet-complete theories where the two components of seesaw arise in a natural manner, namely,
	\begin{itemize}
		\item The left-right (LR) symmetric theories of weak interactions based on the gauge group $SU(3)_C \otimes SU(2)_L \otimes SU(2)_R \otimes U(1)_{B-L}$ \cite{Mohapatra:1979ia,Senjanovic:1978ev,Senjanovic:2018xtu,BhupalDev:2018xya,Grimus:1993fx,Mohapatra:1974gc,Pati:1974yy,Senjanovic:1975rk}.
		\item  $SO(10)$ grand-unified theory for all interations \cite{Pernow:2019tuf}.
	\end{itemize}
	The  small neutrino masses that can explain the tiny neutrino mass can also be responsible for the generation of matter-antimatter aymmetry within the model via leptogenesis. Leptogenesis can be explained by the out-of equilibrium decays of the RH neutrinos via the modes $N\rightarrow L_{l}\phi$ and $N\rightarrow L_{l}^c\phi^c$, where, $L_l = (\nu_L l)_{L}^T$ are the $SU(2)_L$ lepton doublets, $\phi$ is the Higgs doublet and the superscript c denotes the CP conjugate. In the presence of CP violation in the Yukawa sector, these decays can lead to lepton asymmetry in the early universe satisfying the three Sakharov conditions. This asymmetry then undergoes thermodynamic evolution with the expansion of the universe and different reactions present in the model have their impact on washing out part of the asymmetry. The remaining final lepton asymmetry is converted to baryon asymmetry via sphaleron processes before the electroweak phase transition.\\
	In the present work, we are working on TeV scale Left-Right Symmetric Model(LRSM), which stands as a suitable BSM framework, which in addition to explaining Baryon Asymmetry of the Universe (BAU) can also put forward acceptable explanations to phenomenology like Neutrinoless Double Beta Decay $(0\nu\beta\beta)$, Lepton Flavor Violation (LFV) etc. \\
The contribution of flavor symmetries in neutrino mass models is indispensable. However, the implementation of flavor symmetry in the model demands the construction of the effective Lagrangian of the model taking into consideration some new/extra particles called flavons hence giving rise to free parameters and additional terms in the Lagrangian. In the current work, instead of using discrete flavor symmetry, we have used modular symmetry which provides the advantage of not requiring the addition of flavons for the realization of the model. Here, we need to focus on the fact that the use of modular symmetry does not restrict the use of infinite number of terms within a particular model, however when considering a modular group say $\Gamma(N)$, if $2 \leq N \leq 5$, the particular modular group is isomorphic to a non-abelian discrete symmetry group corresponding to the level $N$ of the modular group. In the present work, we have used modular group of level 3, $\Gamma(3)$ for realization and it is isomorphic to the non-abelian $A_4$ discrete symmetry group. And using modular symmetry demands the expression of the Yukawa couplings in terms of modular forms, the number depending upon the level and weight of the modular group used as has been discussed in the succeeding sections.\\
	This work however, mainly focuses on the study and analysis of the effect of assigning different modular weights to the associated modular forms. Different weights result in different textures of the resulting light neutrino mass matrix, which in LRSM is expressed as a summation of type-I and type-II seesaw masses. These difference in the neutrino mass texture have direct implications in the phenomenology associated within the model, the details of which have been discussed and analyzed further within the manuscript. The study of neutrino mass textures have attained sufficient attention because there are no conceiveable set of experiments that can put forward satisfactory explanations about the neutrino mass textures. Now, while implementing modular symmetry, we have used weights $k_{Y}=4,6,8$ and $10$ for the realization of the model and as will be seen later, when using weights $4$ and $10$, we attain one of the seven allowed $2-0$ neutrino mass texture, which is generally referred to as \textbf{Class $B_{2}$} of neutrino mass texture, that is, it is the case where for a symmetric mass matrix $M_{e\mu}=M_{\mu e}=0$ and $M_{\tau\tau}=0$. Using weight $k_{Y}=6$ will result in a matrix that has greater than $2-0$ texture which is disallowed by particle physics and cosmology data and for weight $k_{Y}=8$, we get a texture of neutrino mass where no element is zero, as will be observed further.\\ 
This work also focuses on the analysis of impact neutrino mass textures in Neutrinoless Double Beta Decay $(0\nu\beta\beta)$ within the context of LRSM, Resonant Leptogenesis (RL) and we also analyze the results corresponding to Lepton Flavor Violation (LFV). The main aim in the search of $(0\nu\beta\beta)$ is the measurement of effective Majorana neutrino mass, which is a combination of the neutrino mass eigenstates and neutrino mixing matrix terms \cite{Boruah:2021ktk}. However, no experimental evidence regarding the decay has been in picture till date. In addition to the determination of the effective masses, the half-life of the decay \cite{Ge:2017erv} combined with sufficient knowledge of the nuclear matrix elements (NME), we can set a constraint involving the neutrino masses. The experiments like KamLAND-Zen \cite{Shirai:2018ycl} and GERDA \cite{GERDA:2020xhi} which uses Xenon-136 and Germanium-76 respectively have improved the lower bound on the half-life of the decay process. However, KamLAND-Zen imposes the best lower limit on the half life as $T_{1/2}^{0\nu}>1.07\times10^{26}$ yr at 90 percent CL and the corresponding upper limit of the effective Majorana mass in the range (0.061-0.165)eV. There are several contributions in LRSM that appear due to additional RH current interactions, giving rise to sizeable LFV rates for TeV scale RH neutrino that occur at rates accessible in current experiments. It has been found that the most significant constraints has been provided by the decays, $\mu\rightarrow3e$ and $\mu\rightarrow\gamma e$. In the Standard Model, these LFV decays are suppressed by the tiny neutrino masses. No experiment has so far observed any flavor violating processes including charged leptons. However, many experiments are currently going on to set strong limits on the most relevant LFV observables that will constrain the parameter space of many new models. The best bounds on the branching ratio for LFV decays of the form $\mu\rightarrow\gamma e$ comes from MEG experiment and it is set at $BR(\mu\rightarrow\gamma e)< 4.2 \times 10^{-13}$. In case of the decay $\mu\rightarrow 3e$, the bound is set by the SINDRUM experiment at $BR(\mu\rightarrow 3e)< 1.0 \times 10^{-12}$. In the present work, we have used the global $3\sigma$ \cite{Esteban:2020cvm} values for the calculation of the relevant neutrino parameters.\\
	The paper has been organized as follows, in section \ref{lrsm12} we describe the realization of Left-Right Symmetric Model (LRSM) with $A_4$ modular symmetry. In section \ref{lrsm13}, we discuss about resonant leptogenesis (RL), neutrinoless double beta decay $(0\nu\beta\beta)$ and lepton flavor violation (LFV), in \ref{lrsm14}, we discuss the numerical analysis and results obtained in the present work and summarized the work in the conclusion in section \ref{lrsm15} .
	\begin{center}
		\section{\label{lrsm12}\textbf{Left-Right Symmetric Model and $A_4$ modular symmetry}}
	\end{center}
	Left-Right Symmetric Model has been widely studied in literature in the context of neutrino phenomenology \cite{Lee:2017mfg}. In LRSM, both the left-handed and right-handed components transform as doublet under the $SU(2)$ gauge group. The right-handed neutrino is inherently present within the model and it transforms as a doublet under the $SU(2)$ gauge group. The scalar sector of the model consists of a Higgs bidoublet $\phi (1,2,2,0)$ and two scalar triplets, $\Delta_L(1,3,1,2)$ and $\Delta_R(1,1,3,2)$. Because of the presence of the RH neutrino and the scalar triplet, type-I and type-II seesaw masses appear naturally in the model. \\
	Now, the fermions can attain mass when the Higgs bidoublet couples with the particle content of the model giving rise to the necessary Yukawa Lagrangian in the charged and neutral lepton sector and, the coupling of scalar triplets with the particle content of the model gives rise to the Majorana mass to the neutrinos. The Yukawa Lagrangian giving rise to the Dirac and Majorana mass term are given as,
	\begin{equation}
		\label{x1}
		\mathcal{L_{D}} = \overline{l_{iL}}(Y_{ij}^l \phi + \widetilde{Y_{ij}^l}\widetilde{\phi})l_{jR}+ h.c   
	\end{equation}
	\begin{equation}
		\label{x2}
		\mathcal{L_M}=f_{L,ij}{\Psi_{L,i}}^TCi\sigma_2\Delta_L\Psi_{L,j}+f_{R,ij}{\Psi_{R,i}}^TCi\sigma_2\Delta_R\Psi_{R,j}+h.c
	\end{equation},
	where, $l_L$ and $l_R$ are the left-handed and right-handed lepton fields. $Y^l$ being the Yukawa coupling corresponding to leptons. The Majorana Yukawa couplings $f_L$ and $f_R$ are equal because of the discrete left-right symmetry. The family indices $i,j$ runs from $1$ to $3$ representing the three generations of the fermions. $C=i\gamma_{2}\gamma_{0}$ is the charge conjugation operator, where $\gamma_{\mu}$ are the Dirac matrices and  $\tilde{\phi} = \tau_{2}\phi^{*}\tau_{2}$.\\The symmetry breaking of the gauge group describing the model takes place in two steps. The model gauge group is first broken down to the Standard Model gauge group by the VEV of the scalar triplet $\Delta_R$, and then the Standard Model gauge group is broken down to the electromagnetic gauge group i.e., $U(1)_{em}$ by the VEV of the bidoublet and a tiny vev of the scalar triplet $\Delta_L$.\\So the resultant light neutrino mass of LRSM is expressed as a sum of the type-I and type-II seesaw mass terms, given as,
	\begin{equation}
		\label{x3}
		M_{\nu} = M_{\nu}^I + M_{\nu}^{II}
	\end{equation}
	where,
	\begin{equation}
		\label{x4}
		M_{\nu}^I = M_{D}M_{R}^{-1}M_{D}^T,  	M_{\nu}^{II} = M_{LL}
	\end{equation}

	where,$M_{D}$ is the Dirac mass matrix and $M_{R}$ is the Majorana mass dependent upon the vev of the scalar triplet as $M_{R} = \sqrt{2}v_{R}f_{R}$ and $M_{LL} =  \sqrt{2}v_{L}f_{L}$. Owing to the left-right symmetry, in LRSM the Majorana Yukawa couplings are equal, that is, $f_{L} = f_{R}$. The magnitudes of the VEVs follows the relation, $|v_L|^2 < |k^{2} +k'^{2}| < |v_R|^2$, where $v = \sqrt{k^2 + k'^2}$ is the vev of the Higgs bidoublet $\phi$. In LRSM however, the type-I and type-II mass terms can be expressed in terms of the heavy right-handed Majorana mass matrix, so Eq.\eqref{x3} will follow,
	\begin{equation}
		\label{x8}
		M_\nu = M_D M_{R}^{-1} M_D^T + \gamma\Biggl(\frac{M_W}{v_R}\Biggl)^2 M_{RR}
	\end{equation}
	where, $\gamma$ is a dimensionless parameter which is a function of various couplings, appearing in the VEV of the triplet Higgs $\Delta_L$, i.e., $v_L = \gamma (\frac{v^2}{v_R})$ 
	\begin{equation}
		\label{x9}
		\gamma =  \frac{\beta_1 k k' + \beta_2 k^2 + \beta_3 k'^2}{(2\rho_1 - \rho_3)(k^2 + k'^2)}
	\end{equation} 
	In our model, the dimensionless parameter $\gamma$ has been fine tuned to $\gamma \approx 10^{-15}$ and $v_R$ is considered to be $10 TeV$. $\beta_i$ and $\rho_i$ are dimensionless parameters which appear in the potential of the model.\\
	The light neutrino mass obtained can be expressed in terms of a matrix given as,
	\begin{equation}
		\label{x10}
		M_{\nu}=\begin{pmatrix}
			M_{LL} & M_{D}\\
			M_{D}^{T} & M_{RR}
		\end{pmatrix}
	\end{equation}
	This matrix is a $6 \times 6$ matrix which can be diagonalized by a unitary matrix as follows,
	\begin{equation}
		\label{x11}
		\nu^{T}M_{\nu}\nu = \begin{pmatrix}
			\hat M_{\nu} & 0\\
			0 & \hat M_{RR}
		\end{pmatrix}
	\end{equation}
	where, $\nu$ represents the diagonalizing matrix of the full neutrino mass matrix, $M_{\nu}$,$\hat{M_{\nu}} = diag(m_1,m_2,m_3)$, with $m_i$ being the light neutrino masses and $\hat{M_{RR}} = diag(M_1,M_2,M_3)$, with $M_i$ being the heavy right-handed neutrino masses.\\
	The diagonalizing matrix can be represented as,\\
	\begin{equation}
		\label{x12}
		\nu = \begin{pmatrix}
			U & S\\
			T & V
		\end{pmatrix} \approx \begin{pmatrix}
			1-\frac{1}{2}RR^\dagger & R\\
			-R^\dagger & 1-\frac{1}{2}R^\dagger R
		\end{pmatrix} \begin{pmatrix}
			V_{\nu} & 0\\
			0 & V_R
		\end{pmatrix}
	\end{equation}
	where, $R$ describes the left-right mixing and is given by,\\
	\begin{equation}
		\label{x13}
		R = M_{D}M_{RR}^{-1} + O(M_{D}^3(M_{RR}^{-1})).
	\end{equation}
	The matrices $U,V,S$ and $T$ are as follows,
	\begin{equation}
		\label{x14}
		U = [1-\frac{1}{2}M_{D}M_{RR}^{-1}(M_D M_{RR}^{-1})^\dagger]V_{\nu}
	\end{equation}
	\begin{equation}
		\label{x15}
		V = [1-\frac{1}{2}(M_{D}M_{RR}^{-1})^{\dagger} M_D M_{RR}^{-1}]V_{\nu}
	\end{equation}
	\begin{equation}
		\label{x16}
		S = M_D M_{RR}^{-1} v_{R} f_{R}
	\end{equation}
	\begin{equation}
		\label{x17}
		T = -(M_D M_{RR}^{-1})^\dagger V_{\nu}
	\end{equation} In this work we use modular symmetry for the realization of the model. Several works have already been done on model building and different phenomenology using modular symmetry \cite{Kakoti:2023isn,Kakoti:2023xkn,Kashav:2021zir,CentellesChulia:2023osj,Gogoi:2022jwf}.\\
	Now, after the incorporation of $A_4$ modular symmetry \cite{Feruglio:2017spp},\cite{Novichkov:2019sqv} into the model, the Yukawa Lagrangian takes the form \cite{Kakoti:2023isn,Kakoti:2023xkn},
	\begin{equation}
		\label{x18}
		\mathcal{L_Y} = \overline{l_L}\phi{l_R}Y+\overline{l_L}\tilde{\phi}{l_R}Y+{{l_R}^T}C i{\tau_2}{\Delta_R}{l_R}{Y}+{{l_L}^T}C i{\tau_2}{\Delta_L}{l_L}{Y}
	\end{equation}
	As it is evident from the Lagrangian that the Yukawa couplings have been expressed in terms of the modular form $Y$. When we consider modular group of level 3, that is $\Gamma(3)$, the said group is isomorphic to non-abelian discrete symmetry group $A_{4}$ and the number of modular forms will depend upon the weight of the group under consideration. The number of modular forms required for the construction of a model under modular symmetry is given in table \ref{table:1}.
	\begin{table}[H]
		\begin{center}
			\begin{tabular}{|c|c|c|}
				\hline
				N & No. of modular forms & $\Gamma(N)$ \\
				\hline
				2 & k + 1 & $S_3$ \\
				\hline
				3 & 2k + 1 & $A_4$ \\
				\hline
				4 & 4k + 1 & $S_4$ \\
				\hline
				5 & 10k + 1 & $A_5$ \\
				\hline 
				6 & 12k &  \\
				\hline
				7 & 28k - 2 & \\
				\hline
			\end{tabular}
			\caption{\label{table:1}No. of modular forms corresponding to modular weight 2k.}
		\end{center}
	\end{table}
	The charge assignments for the particle content of the model with incorporation of modular symmetry is given in table \ref{table:2}.
	\begin{table}[H]
		\begin{center}
			\begin{tabular}{|c|c|c|c|c|c|c|}
				\hline
				Gauge group & $l_L$ & $l_R$ & $\phi$ & $\Delta_L$ & $\Delta_R$ \\
				\hline
				$SU(3)_C$ & 1 & 1 & 1 & 1 & 1\\
				\hline
				$SU(2)_L$ & 2 & 1 & 2 & 3 & 1 \\
				\hline
				$SU(2)_R$ & 1 & 2 & 2 & 1 & 3 \\
				\hline
				$U(1)_{B-L}$ & -1 & -1 & 0 & 2 & 2 \\
				\hline 
			\end{tabular}
			\caption{\label{table:2}Charge assignments for the particle content of the model.}
		\end{center}
	\end{table}

Table \ref{table:2} shows the charge assignment for particles within the model under the respective gauge groups and we are working on the basis where our charged lepton mass matrix is diagonal.\\
Using modular symmetry for the realization of a model requires the assignment of particular weights to each of the particle content of the model. As we are concerned about the origin of texture zeros in the resulting light neutrino mass matrix $M_{\nu}$, we will be assigning different weights to $\Gamma(3)$ modular group \cite{Zhang:2019ngf}, which will be having different irreducible representations and hence will help us in achieving the 2-0 textures within the framework under study.
	\begin{itemize}
		\item Considering the weight of the modular group as 2, that is, $k_{Y}=2$, will give us a triplet representation of $Y$ that is, there will be three number of modular forms given as,
		\begin{equation}
			\label{x19}
			Y_{(3)}^{2} = \begin{pmatrix}
				Y_1 \\
				Y_2 \\
				Y_3
			\end{pmatrix}
		\end{equation}
		where, the subscript within brackets represents the multiplet and the superscript stands for the corresponding weight under $A_4$.\\
		\item For $k_{Y}=4$, there will be five number of modular forms, two singlets $1$, $1'$ and one triplet $3$ under $A_{4}$, represented as,
		\begin{equation}
			\label{x20}
			Y_{(1)}^{4} = Y_{1}^{2} + 2 Y_{2} Y_{3}
		\end{equation}
		\begin{equation}
			\label{x21}
			Y_{(1')}^{4} = Y_{3}^{2} + 2 Y_{1} Y_{2}
		\end{equation}
		\begin{equation}
			\label{x22}
			Y_{(3)}^{4} = \begin{pmatrix}
				Y_{1}^{2}-Y_{2}Y_{3}\\
				Y_{3}^{2}-Y_{1}Y_{2}\\
				Y_{2}^{2}-Y_{1}Y_{3}
			\end{pmatrix}
		\end{equation}
		\item For $k_{Y}=6$, there will be seven number of modular forms, two triplets $3_{1}$, $3_{2}$ and one singlet $1$ under $A_{4}$, represented as,
		\begin{equation}
			\label{x23}
			Y_{(1)}^{6} = Y_{1}^{3}+Y_{2}^{3}+Y_{3}^{3}-3Y_{1}Y_{2}Y_{3}
		\end{equation}
		\begin{equation}
			\label{x24}
			Y_{(3_{1})}^{6} = Y_{1}^{2}+2Y_{2}Y_{3}\begin{pmatrix}
				Y_{1}\\
				Y_{2}\\
				Y_{3}
			\end{pmatrix}
		\end{equation}
		\begin{equation}
			\label{x25}
			Y_{(3_{2})}^{6} = Y_{3}^{2}+2Y_{1}Y_{2}\begin{pmatrix}
				Y_{3}\\
				Y_{1}\\
				Y_{2}
			\end{pmatrix}
		\end{equation}
		\item For $k_{Y}=8$, there will be nine number of modular forms, three singlets $1$,$1'$,$1''$ and two triplets $3_{1}$, $3_{2}$ under $A_{4}$, represented as,
		\begin{equation}
			\label{x26}
			Y_{(1)}^{8} = (Y_{1}^{2}+2Y_{2}Y_{3})^{2}
		\end{equation}
		\begin{equation}
			\label{x27}
			Y_{(1')}^{8} = (Y_{1}^{2}+2Y_{2}Y_{3})(Y_{3}^{2}+2Y_{1}Y_{2})
		\end{equation}
		\begin{equation}
			\label{x28}
			Y_{(1'')}^{8} = (Y_{3}^{2}+2Y_{1}Y_{3})^{2}
		\end{equation}
		\begin{equation}
			\label{x29}
			Y_{(3_{1})}^{8} = (Y_{1}^{2}+2Y_{2}Y_{3})\begin{pmatrix}
				Y_{1}^{2}-Y_{2}Y_{3}\\
				Y_{3}^{2}-Y_{1}Y_{2}\\
				Y_{2}^{2}-Y_{1}Y_{3}
			\end{pmatrix}
		\end{equation}
		\begin{equation}
			\label{x30}
			Y_{(3_{2})}^{8} = (Y_{3}^{2}+2Y_{1}Y_{2})\begin{pmatrix}
				Y_{2}^{2}-Y_{1}Y_{3}\\
				Y_{1}^{2}-Y_{2}Y_{3}\\
				Y_{3}^{2}-Y_{1}Y_{2}
			\end{pmatrix}
		\end{equation}
		\item For $k_{Y}=10$, there will be nine number of modular forms, three singlets $1$,$1'$ and three triplets $3_{1}$, $3_{2}$ and $3_{3}$ under $A_{4}$, represented as,
		\begin{equation}
			\label{x31}
			Y_{(1)}^{10} = (Y_{1}^{2}+2Y_{2}Y_{3})(Y_{1}^{3}+Y_{2}^{3}+Y_{3}^{2}-3Y_{1}Y_{2}Y_{3})
		\end{equation}
		\begin{equation}
			\label{x32}
			Y_{(1')}^{10} = (Y_{3}^{2}+2Y_{1}Y_{2})(Y_{1}^{3}+Y_{2}^{3}+Y_{3}^{2}-3Y_{1}Y_{2}Y_{3})
		\end{equation}
		\begin{equation}
			\label{x33}
			Y_{(3_{1})}^{10} = (Y_{1}^{2}+2Y_{2}Y_{3})^{2}\begin{pmatrix}
				Y_{1}\\
				Y_{2}\\
				Y_{3}
			\end{pmatrix}
		\end{equation}
		\begin{equation}
			\label{x34}
			Y_{(3_{2})}^{10} = (Y_{3}^{2}+2Y_{1}Y_{2})^{2}\begin{pmatrix}
				Y_{2}\\
				Y_{3}\\
				Y_{1}
			\end{pmatrix}
		\end{equation}
			\begin{equation}
			\label{x35}
			Y_{(3_{3})}^{10} = (Y_{1}^{2}+2Y_{2}Y_{3})(Y_{3}^{2}+2Y_{1}Y_{2})\begin{pmatrix}
				Y_{3}\\
				Y_{1}\\
				Y_{2}
			\end{pmatrix}
		\end{equation}
	\end{itemize}
	In the current work, we have considered weights $k=4,6,8$ and $10$ for realization of different textures in the neutrino mass matrix, as described below.
	\subsection{\underline{Neutrino mass textures in modular LRSM}}
	\subsubsection{\label{lrsm112}\textbf{Case I - For weight $k_{Y}=4$}}
\begin{figure}[H]
		\centering
		\includegraphics[scale=0.43]{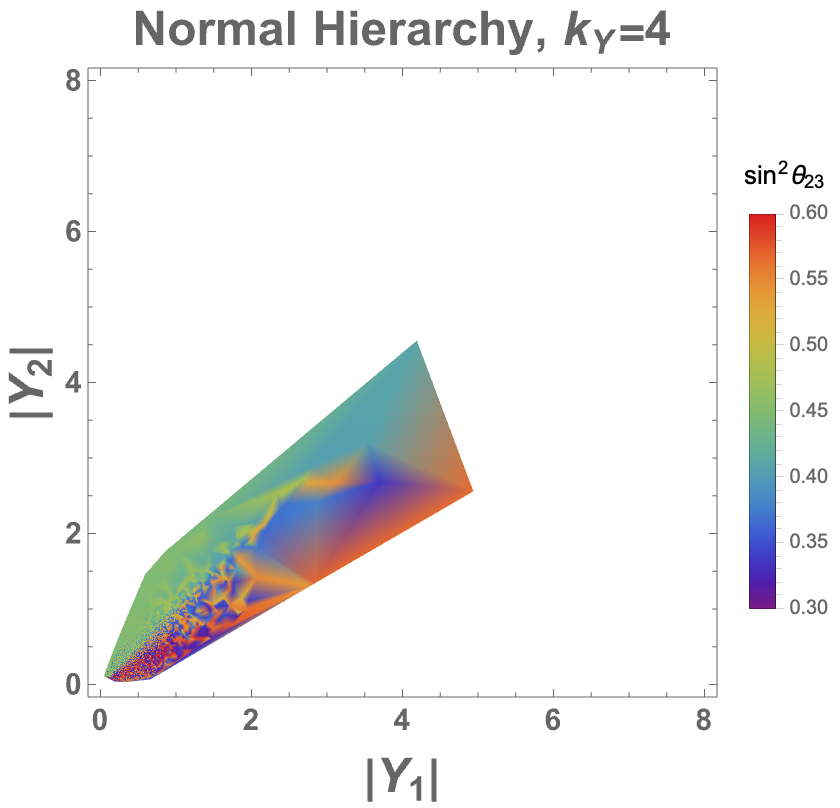}
		\includegraphics[scale=0.43]{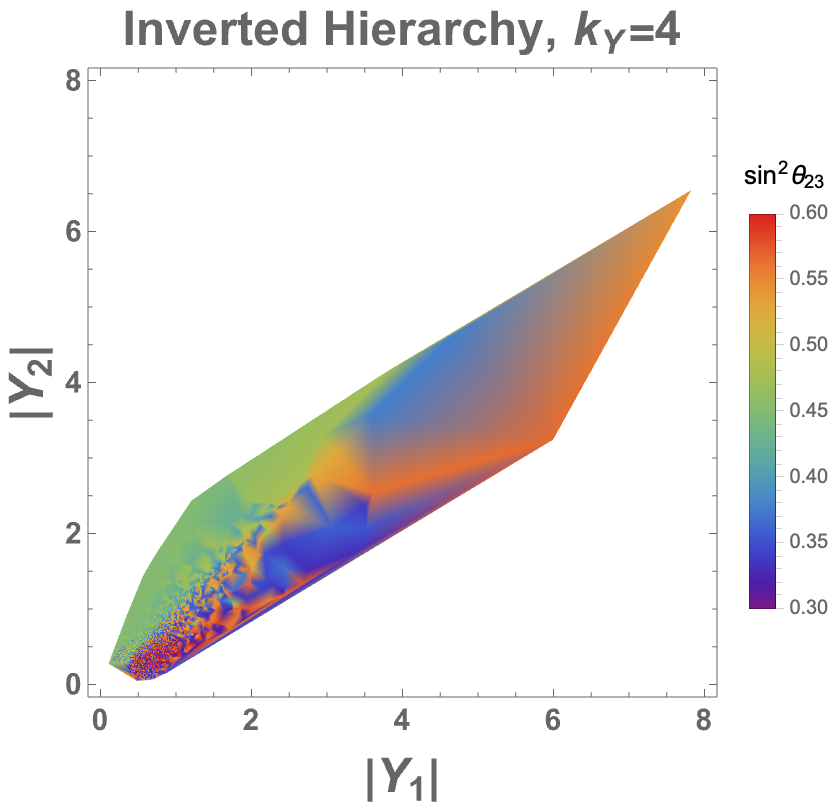}
		\caption{\label{f1} Density plot showing the relation between parameters $|Y_{1}|$, $|Y_{2}|$ and atmospheric mixing angle for normal and inverted hierarchy for weight 4.}
	\end{figure}

\begin{figure}[H]
		\centering
		\includegraphics[scale=0.43]{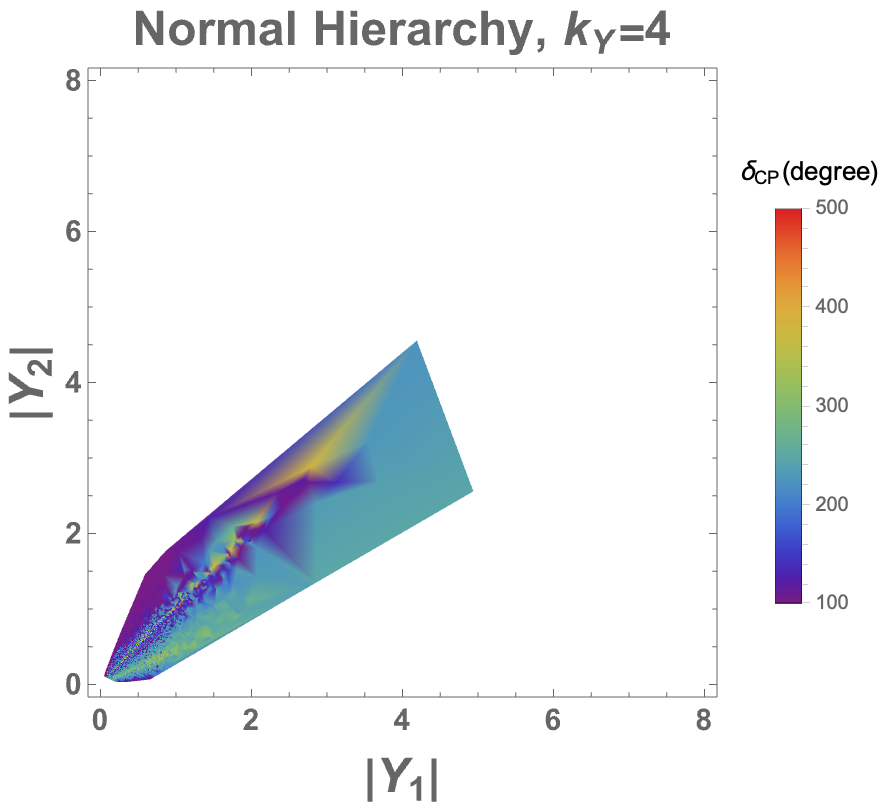}
		\includegraphics[scale=0.43]{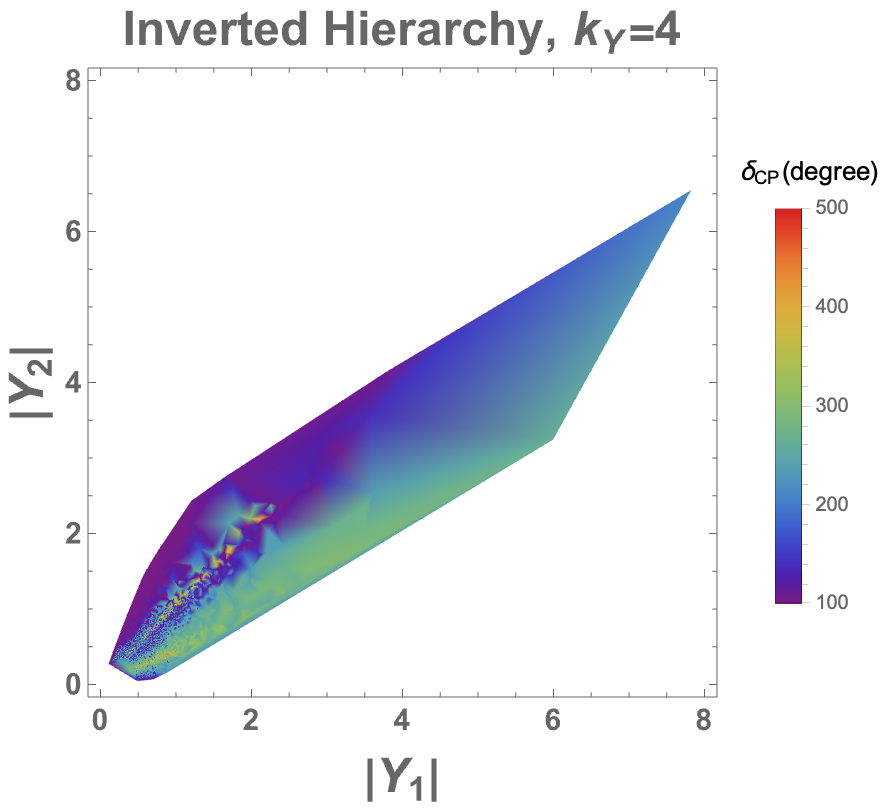}
		\caption{\label{f1a} Density plot showing the relation between parameters $|Y_{1}|$, $|Y_{2}|$ and Dirac CP phase for normal and inverted hierarchy for weight 4.}
	\end{figure}

\begin{figure}[H]
		\centering
		\includegraphics[scale=0.43]{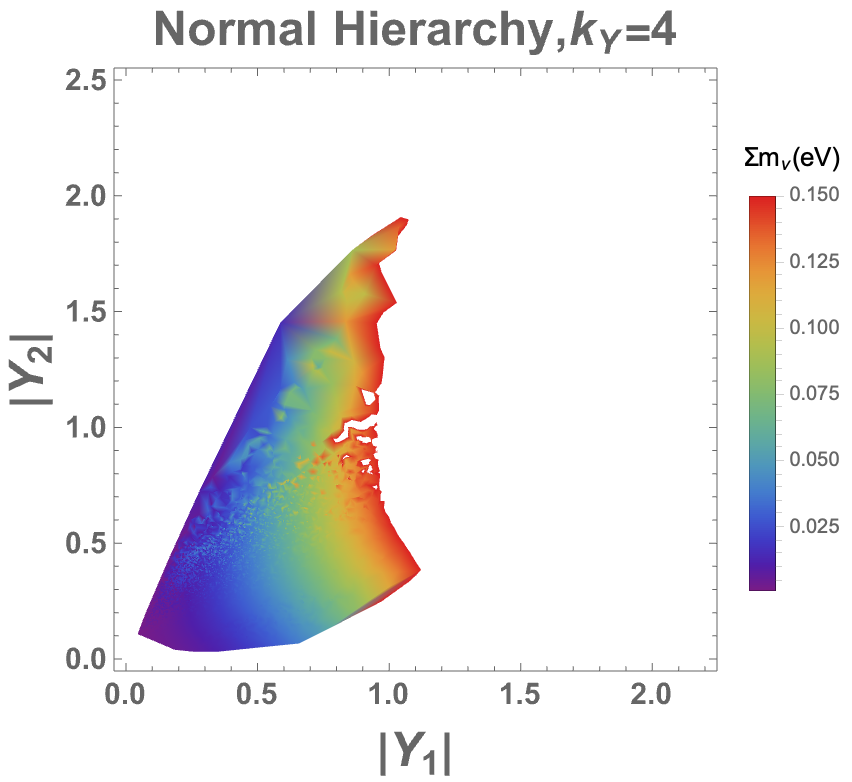}
		\includegraphics[scale=0.43]{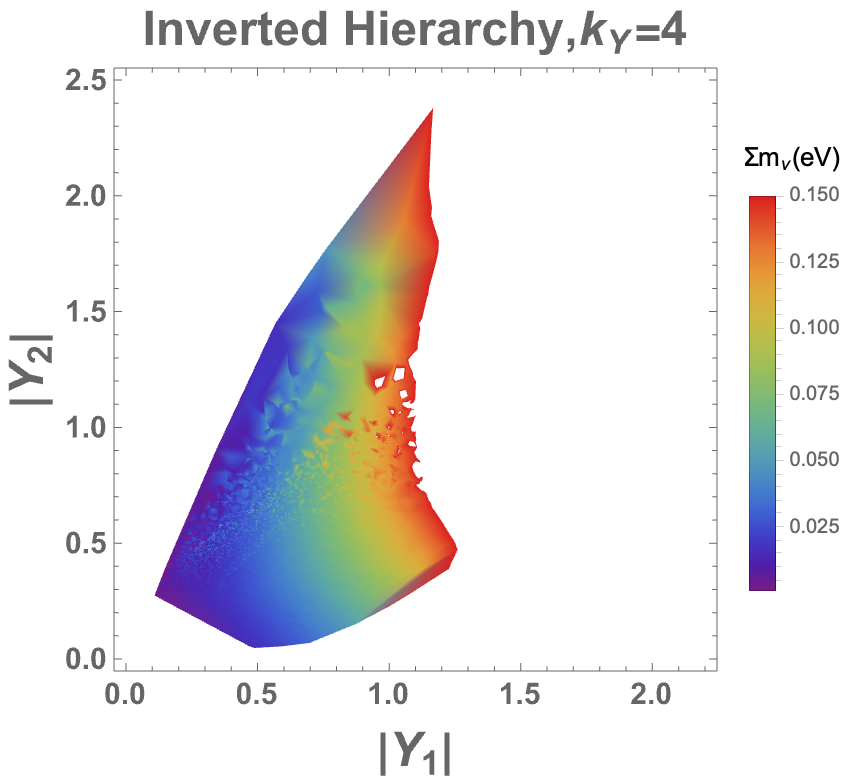}
		\caption{\label{f2} Density plot showing the relation between parameters $|Y_{1}|$, $|Y_{2}|$ and sum of neutrino masses for normal and inverted hierarchy for weight 4.}
	\end{figure}

	As already mentioned in section \ref{lrsm12} for $k_{Y}=4$, we have the singlets $1$ and $1'$, so we consider that the $Y$ transforms as singlets under $A_{4}$. Now, each of the particle within the model will be assigned particular modular weights such that in the Lagrangian, the sum of the modular weights in each term is zero.
		\begin{table}[H]
			\begin{center}
				\begin{tabular}{|c|c|c|c|c|c|c|c|}
					\hline
					& $Y$ & $l_L$ & $l_R$ & $\phi$ & $\Delta_L$ & $\Delta_R$ \\
					\hline
					$A_{4}$ & 1,1' & 1,1',1" & 1,1',1" & 1 & 1 & 1\\
					\hline
					$k$ & 4 & -2 & -2 & -4 & 0 & 0\\
					\hline
				\end{tabular}
				\caption{\label{table:3}Charge and weight assignments for the particle content under $A_4$.}
			\end{center}
		\end{table}
		The Yukawa Lagrangian in equation \eqref{x18} can now be given as,
	 \begin{equation*}
		\label{}
		\mathcal{L_Y} = \phi(\overline{l_{L_{1}}}Y_{(1)}^{4}{l_{R_{1}}}+\overline{l_{L_{1'}}}Y_{(1)}^4{l_{R_{1''}}}+\overline{l_{L_{1''}}}Y_{(1)}^{4}{l_{R_{1'}}}+\overline{l_{L_{1}}}Y_{(1')}^{4}{l_{R_{1''}}}+\overline{l_{L_{1'}}}Y_{(1')}^{4}{l_{R_{1'}}}+\overline{l_{L_{1''}}}Y_{(1')}^{4}{l_{R_{1}}}) +
	\end{equation*}
	\begin{equation*}
		\tilde{\phi}(\overline{l_{L_{1}}}Y_{(1)}^{4}{l_{R_{1}}}+\overline{l_{L_{1'}}}Y_{(1)}^{4}{l_{R_{1''}}}+\overline{l_{L_{1''}}}Y_{(1)}^{4}{l_{R_{1'}}}+\overline{l_{L_{1}}}Y_{(1')}^{4}{l_{R_{1''}}}+\overline{l_{L_{1'}}}Y_{(1')}^{4}{l_{R_{1'}}}+\overline{l_{L_{1''}}}Y_{(1')}^{4}{l_{R_{1}}}) +
	\end{equation*}
	\begin{equation*}
		C i{\tau_2}{\Delta_R}(l_{R_{1}}Y_{(1)}^{4}{l_{R_{1}}}+l_{R_{1'}}Y_{(1)}^{4}{l_{R_{1''}}}+l_{R_{1''}}Y_{(1)}^{4}{l_{R_{1'}}}+l_{R_{1}}Y_{(1')}^{4}{l_{R_{1''}}}+l_{R_{1'}}Y_{(1')}^{4}{l_{R_{1'}}}+l_{R_{1''}}Y_{(1')}^{4}{l_{R_{1}}})+
	\end{equation*}
	\begin{equation}
		\label{x37}
		C i{\tau_2}{\Delta_L}(l_{L_{1}}Y_{(1)}^{4}{l_{L_{1}}}+l_{L_{1'}}Y_{(1)}^{4}{l_{L_{1''}}}+l_{L_{1''}}Y_{(1)}^{4}{l_{L_{1'}}}+l_{L_{1}}Y_{(1')}^{4}{l_{L_{1''}}}+l_{L_{1'}}Y_{(1')}^{4}{l_{L_{1'}}}+l_{L_{1''}}Y_{(1')}^{4}{l_{L_{1}}})
	\end{equation}
		Using $A_{4}$ multiplication rules, we can now determine the Dirac and Majorana mass matrices and hence the resulting light neutrino mass matrix. The Dirac mass matrix is given as,
		\begin{equation}
			M_{D} = v \begin{pmatrix}
				Y_{(1)}^{4} & 0 & Y_{(1')}^{4}\\
				0 & Y_{(1')}^{4} & Y_{(1)}^{4}\\
				Y_{(1')}^{4} & Y_{(1)}^{4} & 0
			\end{pmatrix}
		\end{equation}
		The Majorana mass matrices are given by,
		\begin{equation}
			M_{R} = v_{R} \begin{pmatrix}
				Y_{(1)}^{4} & 0 & Y_{(1')}^{4}\\
				0 & Y_{(1')}^{4} & Y_{(1)}^{4}\\
				Y_{(1')}^{4} & Y_{(1)}^{4} & 0
			\end{pmatrix}
		\end{equation}
		\begin{equation}
			M_{LL} = v_{L} \begin{pmatrix}
				Y_{(1)}^{4} & 0 & Y_{(1')}^{4}\\
				0 & Y_{(1')}^{4} & Y_{(1)}^{4}\\
				Y_{(1')}^{4} & Y_{(1)}^{4} & 0
			\end{pmatrix}
		\end{equation}
		where, $v$, $v_{R}$  and $v_{L}$ are the vevs of the Higgs bidoublet $\phi$ and scalar triplets $\Delta_R$ and $\Delta_L$ respectively. From equations \ref{x20} and \ref{x21}, the terms $Y_{(1)}^4$ and $Y_{(1')}^4$ can be expressed in terms of $(Y_1,Y_2,Y_3)$. So, the resulting light neutrino mass arising as a result of summation of  type-I and type-II seesaw mass can be expressed as,
		\begin{equation}
			\label{x38}
			M_\nu = \begin{pmatrix}
				\frac{(v^{2}+v_{L}v_{R})(Y_{1}^{2}+2Y_{2}Y_{3})}{v_R} & 0 & 	\frac{(v^{2}+v_{L}v_{R})(Y_{3}^{2}+2Y_{1}Y_{2})}{v_R} \\
				0 & \frac{(v^{2}+v_{L}v_{R})(Y_{3}^{2}+2Y_{1}Y_{2})}{v_R} & \frac{(v^{2}+v_{L}v_{R})(Y_{1}^{2}+2Y_{2}Y_{3})}{v_R} \\
				\frac{(v^{2}+v_{L}v_{R})(Y_{3}^{2}+2Y_{1}Y_{2})}{v_R} & \frac{(v^{2}+v_{L}v_{R})(Y_{1}^{2}+2Y_{2}Y_{3})}{v_R} & 0 
			\end{pmatrix}
		\end{equation}
		Here, it has been seen that in \textbf{$M_{\nu}$, $M_{e\mu}=M_{\mu e}=0$} and \textbf{$M_{\tau\tau}=0$}, which is \textbf{Class $B_{2}$} of 2-0 texture as stated in section \ref{lrsm11}. So, for $k=4$, there arises 2-0 texture in the resulting light neutrino mass matrix. After determination of $M_{\nu}$, we calculate the values of $Y_1,Y_2,Y_3$ and then we determine the sum of neutrino masses and go on for further phenomenological studies within the model. Figures \ref{f1} to \ref{f2} show the density plot for the variation of the Yukawa couplings $Y_1,Y_2$ with the neutrino parameters for both the mass orderings. It is evident from the figures that there is  a sharp distinction in the results of both the mass orderings as far as the range of Yukawa couplings are concerned. For normal hierarchy (NH), the value of $Y_1,Y_2$ reaches upto 5 whereas for inverted hierarchy it ranges upto 8. In figure \ref{f1} most of the parameter space is satisfied by the first quadrant of atmospheric mixing angle. In  figure \ref{f1a}, for the variation of $\delta_{CP}$ with Yukawa couplings, the high  density of blue colour signifies the range of $\delta_{CP}$ within (100-200) degrees. Figure  \ref{f2} shows the values of Yukawa couplings that satisfies the neutrino mass sum within the Planck 2018 limit and it is seen that $Y_1$ ranges upto 1 (for both NH and IH) and $Y_2$ has values upto 2 for NH and 2.5 for IH.  
	\subsubsection{\label{lrsm113}\textbf{Case II - For weight $k_{Y}=6$}}
For weight 6, we have seven number of modular forms consisting of one singlet and two triplets given by the expressions \eqref{x23}, \eqref{x24} and \eqref{x25}. The charge and weight assignments for this case are given by,
		\begin{table}[H]
			\begin{center}
				\begin{tabular}{|c|c|c|c|c|c|c|c|}
					\hline
					& $Y$ & $l_L$ & $l_R$ & $\phi$ & $\Delta_L$ & $\Delta_R$ \\
					\hline
					$A_{4}$ & 1& 1,1',1" & 1,1',1" & 1 & 1 & 1\\
					\hline
					$k$ & 6 & -3 & -3 & -6 & 0 & 0\\
					\hline
				\end{tabular}
				\caption{\label{table:4}Charge and weight assignments for the particle content under $A_4$ for k=6.}
			\end{center}
		\end{table}
		The Yukawa Lagrangian in equation \eqref{x18} can now be given as,
		\begin{equation*}
			\label{x39}
			\mathcal{L_Y} = \phi(\overline{l_{L_{1}}}Y_{(1)}^{6}{l_{R_{1}}}+\overline{l_{L_{1'}}}Y_{(1)}^6{l_{R_{1''}}}+\overline{l_{L_{1''}}}Y_{(1)}^{6}{l_{R_{1'}}}+\tilde{\phi}(\overline{l_{L_{1}}}Y_{(1)}^{6}{l_{R_{1}}}+\overline{l_{L_{1'}}}Y_{(1)}^{6}{l_{R_{1''}}}+\overline{l_{L_{1''}}}Y_{(1)}^{6}{l_{R_{1'}}})+
		\end{equation*}
		\begin{equation}
			C i{\tau_2}{\Delta_R}(l_{R_{1}}Y_{(1)}^{6}{l_{R_{1}}}+l_{R_{1'}}Y_{(1)}^{6}{l_{R_{1''}}}+l_{R_{1''}}Y_{(1)}^{6}{l_{R_{1'}}})+C i{\tau_2}{\Delta_L}(l_{L_{1}}Y_{(1)}^{6}{l_{L_{1}}}+l_{L_{1'}}Y_{(1)}^{6}{l_{L_{1''}}}+l_{L_{1''}}Y_{(1)}^{6}{l_{L_{1'}}}
		\end{equation}
		The Dirac and Majorana mass matrices are given as,
		\begin{equation}
			M_{D} = v \begin{pmatrix}
				Y_{(1)}^{6} & 0 & 0\\
				0 & 0 & Y_{(1)}^{6}\\
				0 & Y_{(1)}^{6} & 0
			\end{pmatrix}
		\end{equation}
		\begin{equation}
			M_{R} = v_{R} \begin{pmatrix}
				Y_{(1)}^{6} & 0 & 0\\
				0 & 0 & Y_{(1)}^{6}\\
				0 & Y_{(1)}^{6} & 0
			\end{pmatrix}
		\end{equation}
		\begin{equation}
			M_{LL} = v_{L} \begin{pmatrix}
				Y_{(1)}^{6} & 0 & 0\\
				0 & 0 & Y_{(1)}^{6}\\
				0 & Y_{(1)}^{6} & 0
			\end{pmatrix}
		\end{equation}
	The resulting neutrino mass matrix can be expressed in terms of $Y_1,Y_2,Y_3$ as,
		\begin{equation}
			\label{x40}
			M_\nu = \begin{pmatrix}
				\frac{(v^{2}+v_{L}v_{R})(Y_{1}^{3}+Y_{2}^{3}+Y_{3}^{3}+3Y_{1}Y_{2}Y_{3})}{v_{R}} & 0 & 0\\
				0 & 0 & \frac{(v^{2}+v_{L}v_{R})(Y_{1}^{3}+Y_{2}^{3}+Y_{3}^{3}+3Y_{1}Y_{2}Y_{3})}{v_{R}}\\
				0 & \frac{(v^{2}+v_{L}v_{R})(Y_{1}^{3}+Y_{2}^{3}+Y_{3}^{3}+3Y_{1}Y_{2}Y_{3})}{v_{R}} & 0
			\end{pmatrix}
		\end{equation}
		Here, $M_{e \mu}=M_{\mu e}=0$, $M_{e\tau}=M_{\tau e}=0$, $M_{\mu\mu}=0$ and $M_{\tau\tau}=0$ which is greater than 2-0 texture and is hence not allowed by the recent neutrino and cosmology data. So, for weight, $k_{Y}=6$ we do not get any 2-0 texture of the neutrino mass matrix.

	\subsubsection{\label{lrsm114}\textbf{Case III - For weight $k_{Y}=8$}}
\begin{figure}[H]
		\centering
		\includegraphics[scale=0.43]{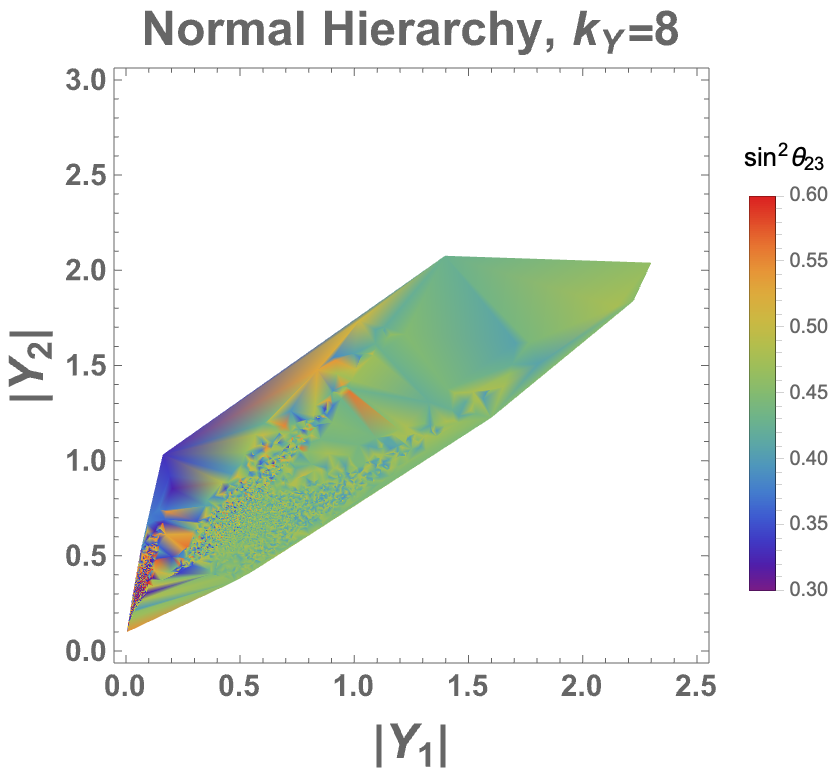}
		\includegraphics[scale=0.43]{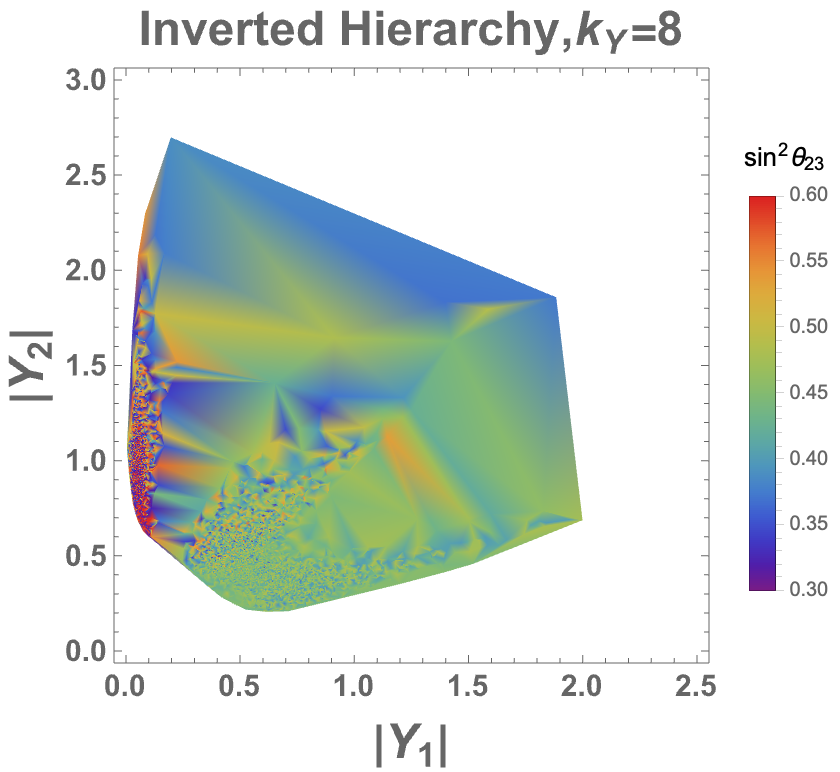}
		\caption{\label{f3}Density plot showing the relation between parameters $|Y_{1}|$, $|Y_{2}|$ and atmospheric mixing angle for normal and inverted hierarchy for weight 8.}
	\end{figure}

\begin{figure}[H]
		\centering
		\includegraphics[scale=0.43]{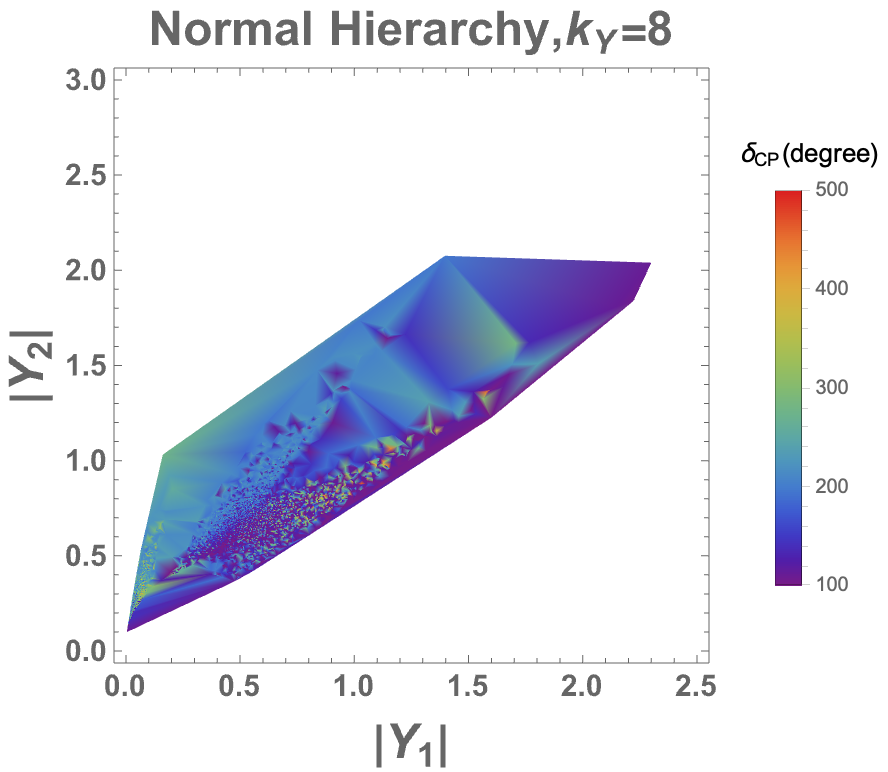}
		\includegraphics[scale=0.43]{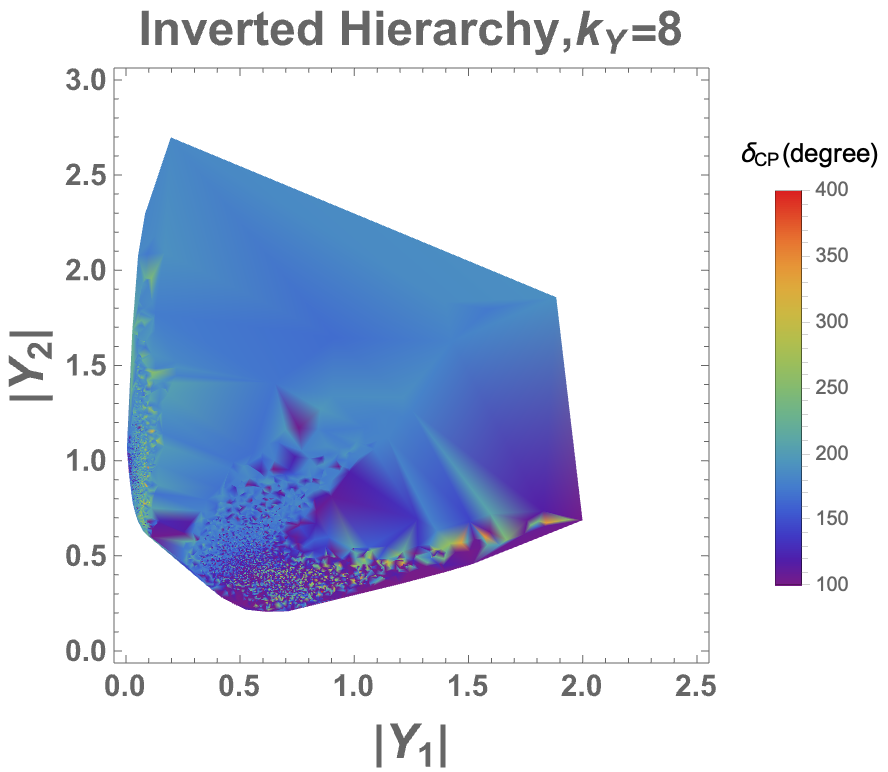}
		\caption{\label{f3a}Density plot showing the relation between parameters $|Y_{1}|$, $|Y_{2}|$ and Dirac CP phase for normal and inverted hierarchy for weight 8.}
	\end{figure}

\begin{figure}[H]
		\centering
		\includegraphics[scale=0.43]{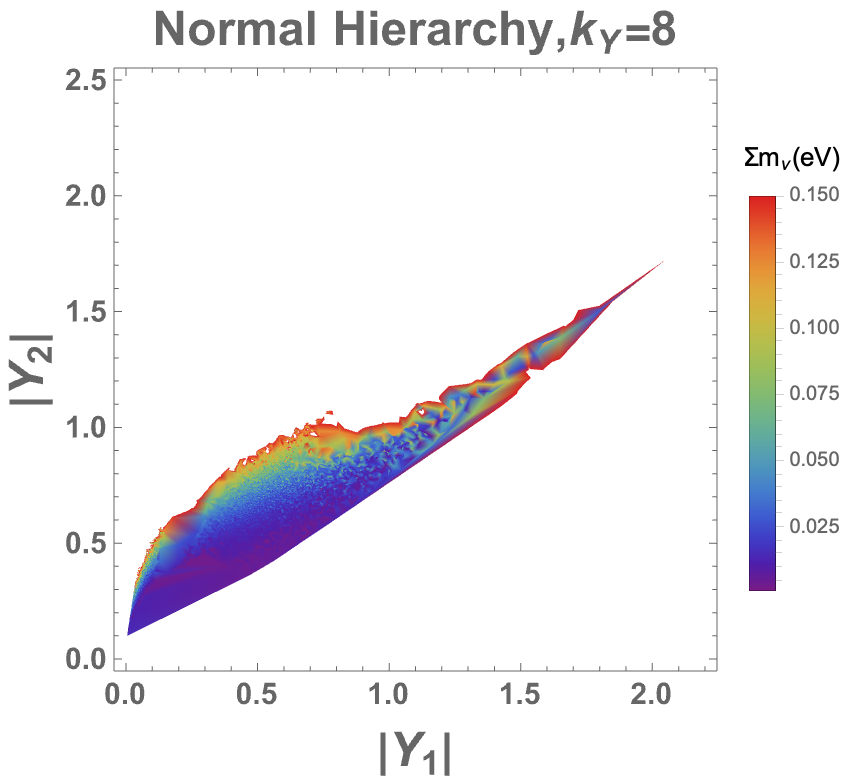}
		\includegraphics[scale=0.43]{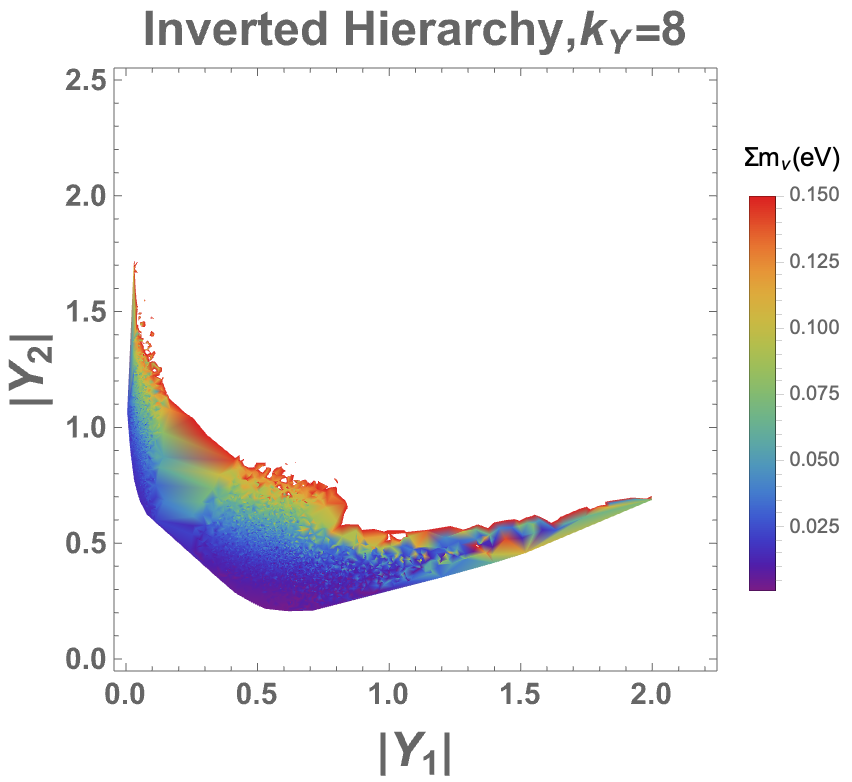}
		\caption{\label{f4} Density plot showing the relation between parameters $|Y_{1}|$, $|Y_{2}|$ and sum of neutrino masses for normal and inverted hierarchy for weight 8.}
	\end{figure}

For weight 8, there are nine modular forms, two triplets and three singlets as shown in equations \ref{x26} to \ref{x29}. Considering the following charge assignments, we determine the Yukawa Lagrangian for the respective weight.
		\begin{table}[H]
			\begin{center}
				\begin{tabular}{|c|c|c|c|c|c|c|c|}
					\hline
					& $Y$ & $l_L$ & $l_R$ & $\phi$ & $\Delta_L$ & $\Delta_R$ \\
					\hline
					$A_{4}$ & 1,1',1"& 1,1',1" & 1,1',1" & 1 & 1 & 1\\
					\hline
					$k$ & 8& -4 & -4 & -8 & 0 & 0\\
					\hline
				\end{tabular}
				\caption{\label{table:5}Charge and weight assignments for the particle content under $A_4$ for weight 8.}
			\end{center}
		\end{table}
		The Yukawa Lagrangian in equation \eqref{x18} can now be given as,
		\begin{equation*}
			\label{x41}
			\mathcal{L_Y} = \phi(\overline{l_{L_{1}}}Y_{(1)}^{8}{l_{R_{1}}}+\overline{l_{L_{1'}}}Y_{(1)}^8{l_{R_{1''}}}+\overline{l_{L_{1''}}}Y_{(1)}^{8}{l_{R_{1'}}}+\overline{l_{L_{1}}}Y_{(1')}^{8}{l_{R_{1''}}}+\overline{l_{L_{1'}}}Y_{(1')}^{8}{l_{R_{1'}}}+\overline{l_{L_{1''}}}Y_{(1')}^{8}{l_{R_{1}}} +
		\end{equation*}
		\begin{equation*}
			\overline{l_{L_{1}}}Y_{(1")}^{8}{l_{R_{1'}}}+\overline{l_{L_{1'}}}Y_{(1")}^8{l_{R_{1}}}+\overline{l_{L_{1''}}}Y_{(1")}^{8}{l_{R_{1"}}})+\tilde{\phi}(\overline{l_{L_{1}}}Y_{(1)}^{8}{l_{R_{1}}}+\overline{l_{L_{1'}}}Y_{(1)}^{8}{l_{R_{1''}}}+\overline{l_{L_{1''}}}Y_{(1)}^{8}{l_{R_{1'}}}+
		\end{equation*}
		\begin{equation*}
		\overline{l_{L_{1}}}Y_{(1')}^{8}{l_{R_{1''}}}+\overline{l_{L_{1'}}}Y_{(1')}^{8}{l_{R_{1'}}}+\overline{l_{L_{1''}}}Y_{(1')}^{8}{l_{R_{1}}} +\overline{l_{L_{1}}}Y_{(1")}^{8}{l_{R_{1'}}}+\overline{l_{L_{1'}}}Y_{(1")}^8{l_{R_{1}}}+\overline{l_{L_{1''}}}Y_{(1")}^{8}{l_{R_{1"}}})
			\end{equation*}
		\begin{equation*}
			C i{\tau_2}{\Delta_R}(l_{R_{1}}Y_{(1)}^{4}{l_{R_{1}}}+l_{R_{1'}}Y_{(1)}^{4}{l_{R_{1''}}}+l_{R_{1''}}Y_{(1)}^{4}{l_{R_{1'}}}+l_{R_{1}}Y_{(1')}^{4}{l_{R_{1''}}}+l_{R_{1'}}Y_{(1')}^{4}{l_{R_{1'}}}+l_{R_{1''}}Y_{(1')}^{4}{l_{R_{1}}}+
		\end{equation*}
		\begin{equation*}
			l_{R_{1}}Y_{(1")}^{8}{l_{R_{1'}}}+l_{R_{1'}}Y_{(1")}^{8}{l_{R_{1}}}+l_{R_{1''}}Y_{(1")}^{8}{l_{R_{1"}}})+C i{\tau_2}{\Delta_L}(l_{L_{1}}Y_{(1)}^{8}{l_{L_{1}}}+l_{L_{1'}}Y_{(1)}^{8}{l_{L_{1''}}}+l_{L_{1''}}Y_{(1)}^{8}{l_{L_{1'}}}
		\end{equation*}
		\begin{equation}
			\label{x42}
			+l_{L_{1}}Y_{(1')}^{8}{l_{L_{1''}}}+l_{L_{1'}}Y_{(1')}^{8}{l_{L_{1'}}}+l_{L_{1''}}Y_{(1')}^{8}{l_{L_{1}}}+l_{L_{1}}Y_{(1")}^{8}{l_{L_{1'}}}+l_{L_{1'}}Y_{(1")}^{8}{l_{L_{1}}}+l_{L_{1"}}Y_{(1")}^{8}{l_{L_{1"}}})
		\end{equation}
		The Dirac and Majorana mass matrices can now be determined as,
		\begin{equation}
			M_{D} = v \begin{pmatrix}
				Y_{(1)}^{8} & Y_{1''}^{8} & Y_{1'}^{8}\\
				Y_{1''}^{8}& Y_{1'}^{8} & Y_{(1)}^{8}\\
				Y_{1'}^{8}& Y_{(1)}^{8} & Y_{1''}^{8}
			\end{pmatrix}
		\end{equation}
		The Majorana mass matrices are given by,
		\begin{equation}
			M_{R} = v_{R} \begin{pmatrix}
				Y_{(1)}^{8} & Y_{1''}^{8} & Y_{1'}^{8}\\
				Y_{1''}^{8}& Y_{1'}^{8} & Y_{(1)}^{8}\\
				Y_{1'}^{8}& Y_{(1)}^{8} & Y_{1''}^{8}
			\end{pmatrix}
		\end{equation}
		\begin{equation}
			M_{LL} = v_{L}\begin{pmatrix}
				Y_{(1)}^{8} & Y_{1''}^{8} & Y_{1'}^{8}\\
				Y_{1''}^{8}& Y_{1'}^{8} & Y_{(1)}^{8}\\
				Y_{1'}^{8}& Y_{(1)}^{8} & Y_{1''}^{8}
			\end{pmatrix}
		\end{equation}
		In terms of $Y_{1},Y_{2},Y_{3}$ , the mass matrix can be written as,
		\begin{equation}
			\label{x43}
			M_\nu = \begin{pmatrix}
				\frac{(v^{2}+v_{L}v_{R})(Y_{1}^{2}+2 Y_{2} Y_{3})^{2}}{v_{R}} & \frac{(v^{2}+v_{L}v_{R})(Y_{3}^{2}+2 Y_{1} Y_{2})^{2}}{v_{R}} & \frac{(v^{2}+v_{L}v_{R})(Y_{1}^{2}+2Y_{2}Y_{3})(2Y_{1}Y_{2}+Y_{3}^{2})}{v_{R}}\\
				\frac{(v^{2}+v_{L}v_{R})(Y_{3}^{2}+2 Y_{1} Y_{2})^{2}}{v_{R}} & \frac{(v^{2}+v_{L}v_{R})(Y_{1}^{2}+2Y_{2}Y_{3})(2Y_{1}Y_{2}+Y_{3}^{2})}{v_{R}} & \frac{(v^{2}+v_{L}v_{R})(Y_{1}^{2}+2 Y_{2} Y_{3})^{2}}{v_{R}}\\
				\frac{(v^{2}+v_{L}v_{R})(Y_{1}^{2}+2Y_{2}Y_{3})(2Y_{1}Y_{2}+Y_{3}^{2})}{v_{R}} &\frac{(v^{2}+v_{L}v_{R})(Y_{1}^{2}+2 Y_{2} Y_{3})^{2}}{v_{R}} & \frac{(v^{2}+v_{L}v_{R})(Y_{3}^{2}+2 Y_{1} Y_{2})^{2}}{v_{R}} 
			\end{pmatrix}
		\end{equation}
	So, equation \eqref{x43} shows that all the elements in the resulting neutrino mass matrix are non-vanishing and as such we have seen the variations of the Yukawa couplings with the neutrino mixing parameters and also with sum of neutrino masses which has been shown in figures \ref{f3} to \ref{f4} in density plots. Again there is a variation in the range of Yukawa couplings in both the mass hierarchies with NH covering less parameter space and in both the case the value of the atmospheric mixing angle satisfying the lower quadrant, value of $\delta_{CP}$ not more than 200 degree.

	\subsubsection{\label{lrsm114a}\textbf{Case III - For weight $k_{Y}=10$}}
		As already mentioned in section \ref{lrsm12} for $k_{Y}=10$, we have the singlets $1$ and $1'$, so we consider that the $Y$ transforms as singlets under $A_{4}$. Now, each of the particle within the model will be assigned particular modular weights such that in the Lagrangian, sum of the modular weights in each term is zero.
	\begin{table}[H]
		\begin{center}
			\begin{tabular}{|c|c|c|c|c|c|c|c|}
				\hline
				& $Y$ & $l_L$ & $l_R$ & $\phi$ & $\Delta_L$ & $\Delta_R$ \\
				\hline
				$A_{4}$ & 1,1' & 1,1',1" & 1,1',1" & 1 & 1 & 1\\
				\hline
				$k$ & 10 & -5 & -5 & -10 & 0 & 0\\
				\hline
			\end{tabular}
			\caption{\label{table:3a}Charge and weight assignments for the particle content under $A_4$.}
		\end{center}
	\end{table}
	The Yukawa Lagrangian in equation \eqref{x18} can now be given as,
	\begin{equation*}
		\label{x44}
		\mathcal{L_Y} = \phi(\overline{l_{L_{1}}}Y_{(1)}^{10}{l_{R_{1}}}+\overline{l_{L_{1'}}}Y_{(1)}^{10}{l_{R_{1''}}}+\overline{l_{L_{1''}}}Y_{(1)}^{10}{l_{R_{1'}}}+\overline{l_{L_{1}}}Y_{(1')}^{10}{l_{R_{1''}}}+\overline{l_{L_{1'}}}Y_{(1')}^{10}{l_{R_{1'}}}+\overline{l_{L_{1''}}}Y_{(1')}^{10}{l_{R_{1}}}) +
	\end{equation*}
	\begin{equation*}
		\tilde{\phi}(\overline{l_{L_{1}}}Y_{(1)}^{10}{l_{R_{1}}}+\overline{l_{L_{1'}}}Y_{(1)}^{10}{l_{R_{1''}}}+\overline{l_{L_{1''}}}Y_{(1)}^{10}{l_{R_{1'}}}+\overline{l_{L_{1}}}Y_{(1')}^{10}{l_{R_{1''}}}+\overline{l_{L_{1'}}}Y_{(1')}^{10}{l_{R_{1'}}}+\overline{l_{L_{1''}}}Y_{(1')}^{10}{l_{R_{1}}}) +
	\end{equation*}
	\begin{equation*}
		C i{\tau_2}{\Delta_R}(l_{R_{1}}Y_{(1)}^{10}{l_{R_{1}}}+l_{R_{1'}}Y_{(1)}^{10}{l_{R_{1''}}}+l_{R_{1''}}Y_{(1)}^{10}{l_{R_{1'}}}+l_{R_{1}}Y_{(1')}^{10}{l_{R_{1''}}}+l_{R_{1'}}Y_{(1')}^{10}{l_{R_{1'}}}+l_{R_{1''}}Y_{(1')}^{10}{l_{R_{1}}})+
	\end{equation*}
	\begin{equation}
		\label{q2}
		C i{\tau_2}{\Delta_L}(l_{L_{1}}Y_{(1)}^{10}{l_{L_{1}}}+l_{L_{1'}}Y_{(1)}^{10}{l_{L_{1''}}}+l_{L_{1''}}Y_{(1)}^{10}{l_{L_{1'}}}+l_{L_{1}}Y_{(1')}^{10}{l_{L_{1''}}}+l_{L_{1'}}Y_{(1')}^{10}{l_{L_{1'}}}+l_{L_{1''}}Y_{(1')}^{10}{l_{L_{1}}})
	\end{equation}
	Using $A_{4}$ multiplication rules, we can now determine the Dirac and Majorana mass matrices and hence the resulting light neutrino mass matrix. The Dirac mass matrix is given as,
	\begin{equation}
		M_{D} = v \begin{pmatrix}
			Y_{(1)}^{10} & 0 & Y_{(1')}^{10}\\
			0 & Y_{(1')}^{10} & Y_{(1)}^{10}\\
			Y_{(1')}^{10} & Y_{(1)}^{10} & 0
		\end{pmatrix}
	\end{equation}
	The Majorana mass matrices are given by,
	\begin{equation}
		M_{R} = v_{R} \begin{pmatrix}
			Y_{(1)}^{10} & 0 & Y_{(1')}^{10}\\
			0 & Y_{(1')}^{10} & Y_{(1)}^{10}\\
			Y_{(1')}^{10} & Y_{(1)}^{10} & 0
		\end{pmatrix}
	\end{equation}
	\begin{equation}
		M_{LL} = v_{L} \begin{pmatrix}
			Y_{(1)}^{10} & 0 & Y_{(1')}^{10}\\
			0 & Y_{(1')}^{10} & Y_{(1)}^{10}\\
			Y_{(1')}^{10} & Y_{(1)}^{10} & 0
		\end{pmatrix}
	\end{equation}
	Now, in terms of $Y_{1}$,$Y_{2}$ and $Y_{3}$, we can express the resulting light neutrino mass matrix as,
	\begin{equation}
		\label{x45}
		M_\nu = a\begin{pmatrix}
			\frac{(Y_{1}^{2}+2Y_{2}Y_{3})(Y_{1}^{3}+Y_{2}^{3}+Y_{3}^{3}-3Y_{1}Y_{2}Y_{3})}{v_R} & 0 & 	\frac{(Y_{3}^{2}+2Y_{1}Y_{2})(Y_{1}^{3}+Y_{2}^{3}+Y_{3}^{3}-3Y_{1}Y_{2}Y_{3})}{v_R} \\
			0 & \frac{(Y_{3}^{2}+2Y_{1}Y_{2})(Y_{1}^{3}+Y_{2}^{3}+Y_{3}^{3}-3Y_{1}Y_{2}Y_{3})}{v_R} & \frac{(Y_{1}^{2}+2Y_{2}Y_{3})(Y_{1}^{3}+Y_{2}^{3}+Y_{3}^{3}-3Y_{1}Y_{2}Y_{3})}{v_R} \\
			\frac{(Y_{3}^{2}+2Y_{1}Y_{2})(Y_{1}^{3}+Y_{2}^{3}+Y_{3}^{3}-3Y_{1}Y_{2}Y_{3})}{v_R} & \frac{(Y_{1}^{2}+2Y_{2}Y_{3})(Y_{1}^{3}+Y_{2}^{3}+Y_{3}^{3}-3Y_{1}Y_{2}Y_{3})}{v_R} & 0 
		\end{pmatrix}
	\end{equation}
	where, $a=v^{2}+v_{L}v_{R}$.
		\begin{figure}[H]
		\centering
		\includegraphics[scale=0.43]{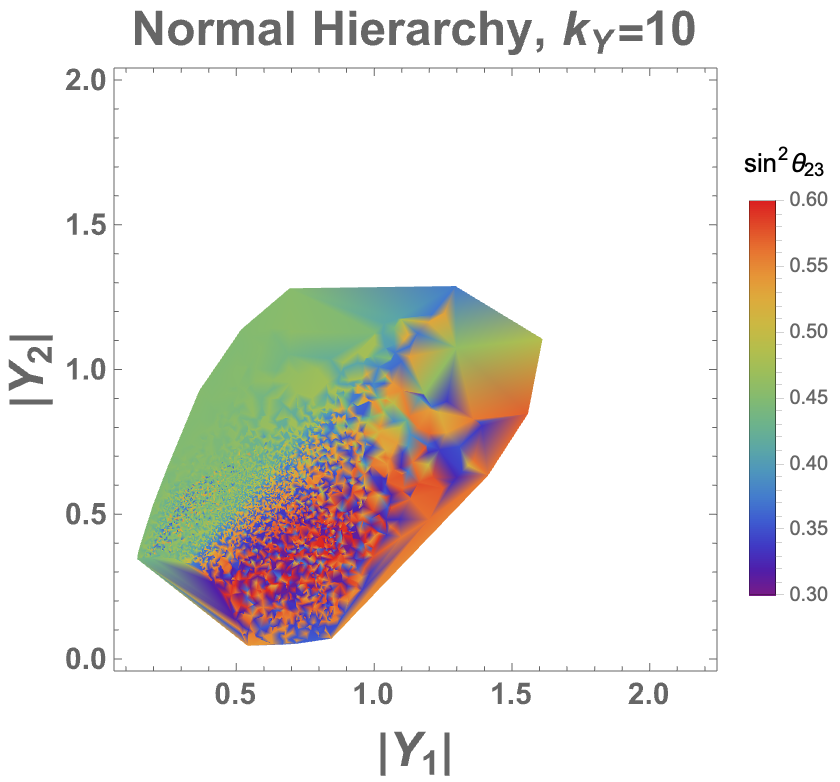}
		\includegraphics[scale=0.43]{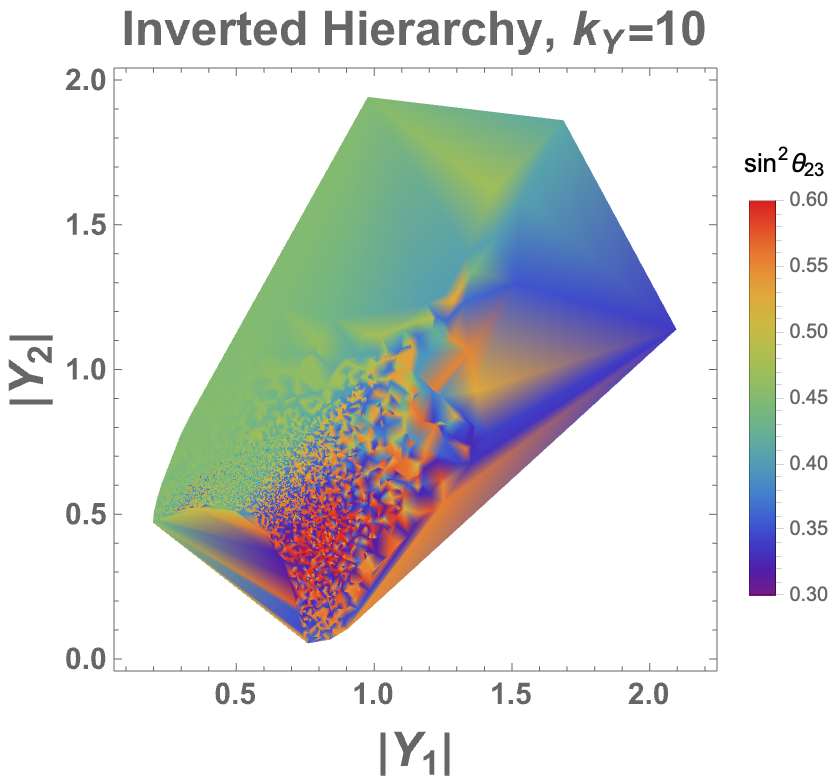}
		\caption{\label{ff3}Density plot showing the relation between parameters $|Y_{1}|$, $|Y_{2}|$ and atmospheric mixing angle for normal and inverted hierarchy for weight 10.}
	\end{figure}
	\begin{figure}[H]
		\centering
		\includegraphics[scale=0.43]{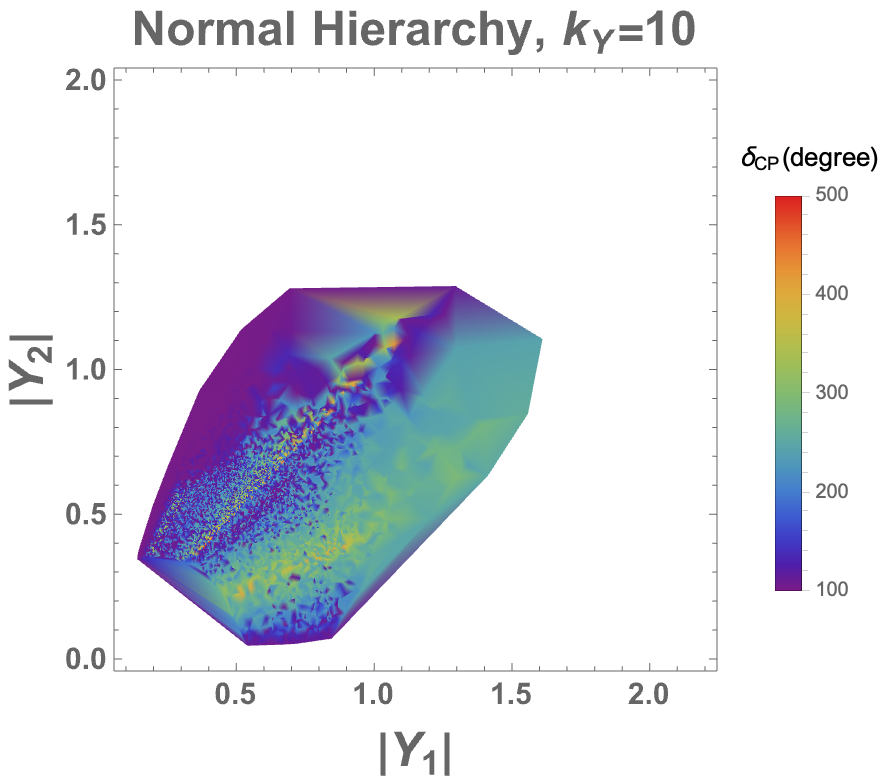}
		\includegraphics[scale=0.43]{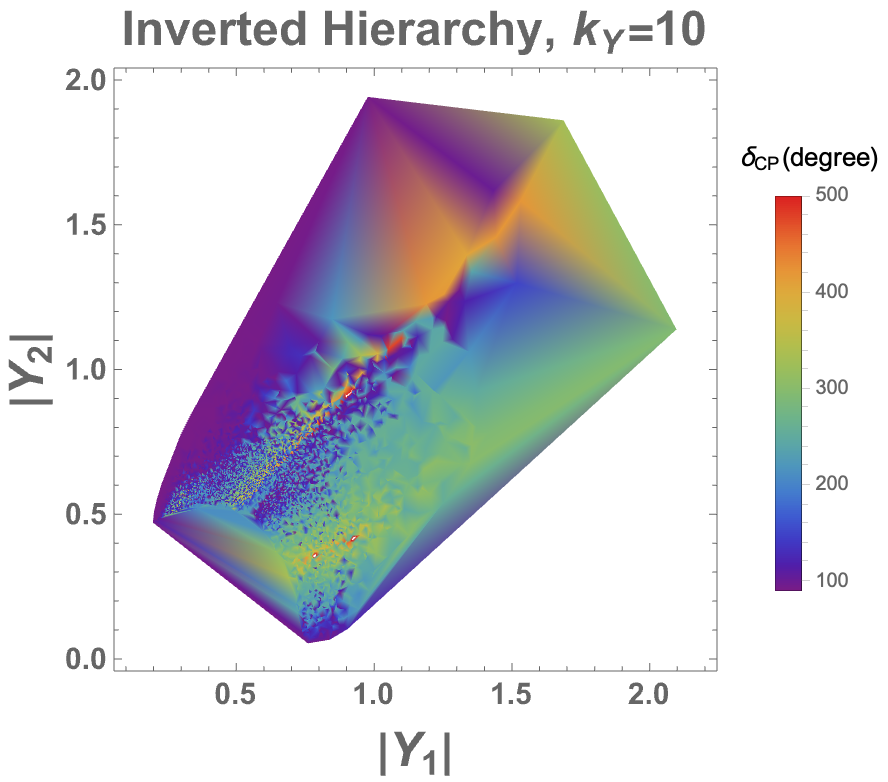}
		\caption{\label{ff4}Density plot showing the relation between parameters $|Y_{1}|$, $|Y_{2}|$ and Dirac CP phase for normal and inverted hierarchy for weight 10.}
	\end{figure}
	\begin{figure}[H]
		\centering
		\includegraphics[scale=0.43]{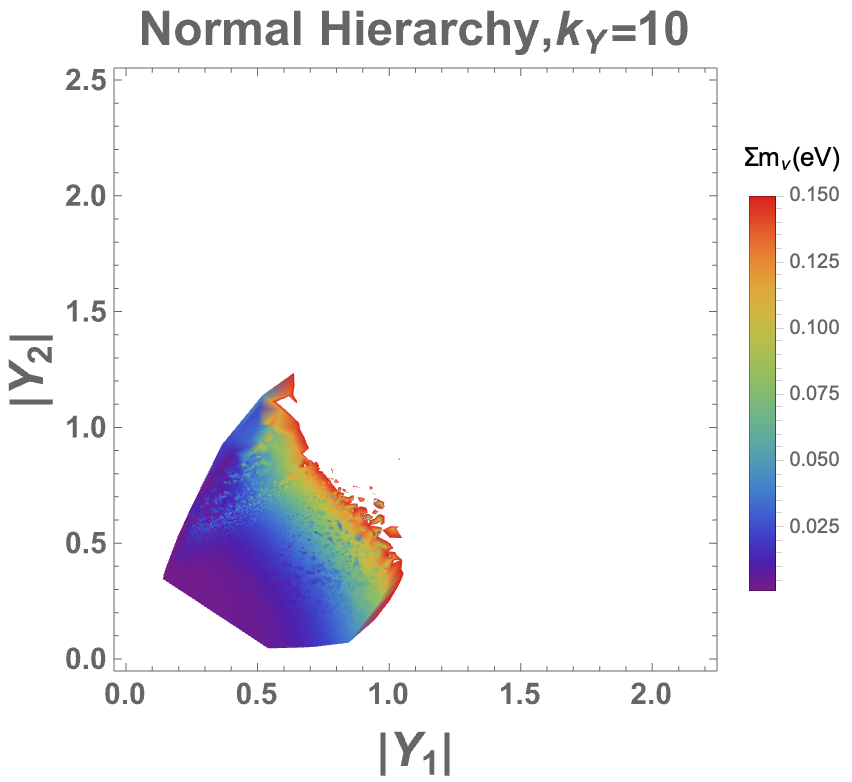}
		\includegraphics[scale=0.43]{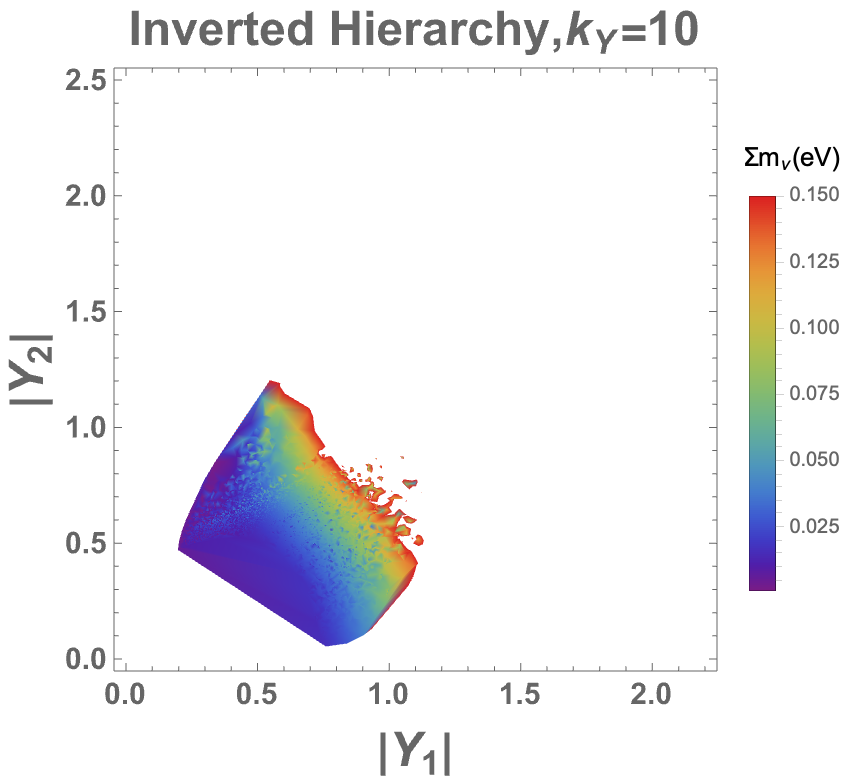}
		\caption{\label{ff5} Density plot showing the relation between parameters $|Y_{1}|$, $|Y_{2}|$ and sum of neutrino masses for normal and inverted hierarchy for weight 10.}
	\end{figure}
	In figures \ref{ff3} to \ref{ff5}, we have shown the density plot for the variation of the neutrino parameters with Yukawa couplings $Y_1,Y_2$ and sum of neutrino mass for both the mass orderings. We see that the range of Yukawa couplings covers a very less parameter space for NH as compared to IH with the value of atmospheric angle preferably in lower quadrant and value of $\delta_{CP}$ less than 250 degree. As far as the neutrino mass sum satisfying PLANCK 2018 bound is concerned, the Yukawa couplings covers very less parameter space irrespective of mass orderings. 

	\begin{center}
		\section{\label{lrsm13}Resonant leptogenesis, lepton number and lepton flavor violation in LRSM}
	\end{center}
	An adequate amount of lepton asymmetry can be generated in TeV scale seesaw models by the phenomenon of resonant leptogenesis (RL) \cite{Asaka:2018hyk,Blanchet:2009bu,Flanz:1996fb,Dev:2015cxa}. The condition for RL is that, two of the right-handed Majorana neutrinos need to be degenerate. As it is known that in LRSM, RH neutrinos and scalar triplets are present, so their decay can give rise to lepton asymmetry, which suggests that the net lepton asymmetry will arise as a result of both the seesaw terms. In the present work, we investigate the validity of texture zero neutrino mass matrix with the phenomenological implication of leptogenesis.\\ The rate at which the RH neutrinos decay is governed by the Yukawa couplings, and is hence given by \cite{Borgohain:2017akh},
	\begin{equation}
		\label{x46}
		\Gamma_{i}=(Y_{\nu}^{\dagger}Y_{\nu})_{ii} \frac{M_i}{8\pi}
	\end{equation}
	One important condition for RL is that the mass difference of the two heavy RH neutrinos must be comparable to their decay width, i.e., $M_{i}-M_{j}=\Gamma$. In such a case, the CP asymmetry may become very large. The CP violating asymmetry is thus given by \cite{Xing:2015fdg},
	\begin{equation}
		\label{x47}
		\epsilon_{i}=\frac{Im[(Y_{\nu}^{\dagger}Y_{\nu})_{ij}^2]}{(Y_{\nu}^{\dagger}Y_{\nu})_{11}(Y_{\nu}^{\dagger}Y_{\nu})_{22}}.\frac{(M_{i}^{2}-M_{j}^{2})M_{i}\Gamma_{j}}{(M_{i}^{2}-M_{j}^{2})^{2}+M_{i}^{2}\Gamma_{j}^{2}}
	\end{equation}
	The variables $i,j$ run over 1 and 2 and $i\neq j$.\\
	The CP asymmetries $\epsilon_{1}$ and $\epsilon_{2}$ can give rise to a net lepton asymmetry, provided the expansion rate of the universe is larger than $\Gamma_{1}$ and $\Gamma_{2}$. This can further be converted into baryon asymmetry of the universe by $B+L$ violating sphaleron processes.\\
	Due to the presence of several new heavy particles in LRSM, along with the standard light neutrino contribution to $0\nu\beta\beta$, several new physics contributions also come into the picture. There are total eight contributions of the phenomenon to LRSM which have been mentioned in \cite{Kakoti:2023isn,Kakoti:2023xkn,Borgohain:2017inp}. However, here we have considered only the right-handed neutrino as well as mixed momentum-dependent contributions for $0\nu\beta\beta$ in this work, namely,
	\begin{itemize}
		\item  Heavy right-handed neutrino contribution where the mediator particles are the $W_R$ bosons. The amplitude of this process is dependent on the elements of the right handed leptonic mixing matrix and mass of the right-handed gauge boson, $W_R$ as well as the mass of the heavy right handed Majorana neutrino.
		\item Light neutrino contribution from the Feynman diagram mediated by both $W_L$ and $W_R$, and the amplitude of the process depends upon the mixing between light and heavy neutrinos, leptonic mixing matrix elements, light neutrino masses and the mass of the gauge bosons, $W_L$ and $W_R$ ($\lambda$ contribution). 
		\item Heavy neutrino contribution from the Feynman diagram mediated by both $W_L$ and $W_R$, and the amplitude of the process depends upon the right handed leptonic mixing matrix elements, mixing between the light and heavy neutrinos, also the mass of the gauge bosons, $W_L$ and $W_R$ and mass of the heavy right handed neutrino ($\eta$ contribution).
	\end{itemize}
	In this work, we calculate the effective masses for both the above mentioned contributions and we study their variations with sum of the neutrino masses and also with the branching ratios corresponding to the lepton flavor violating decays. The calculations and variations with the baryon asymmetry parameter $\eta_B$ has been shown in section \ref{lrsm14}.\\
	Charged Lepton Flavor Violation (CLFV) is a clear signal of new physics, which directly addresses the physics of flavor and generations. The search for this process has been carried out since 1940's when muon was thought to be a separate particle. The LFV effects from new physics at the TeV scale are found to be present in many models and hence is a clear signal of physics BSM \cite{Ardu:2022sbt,Lindner:2016bgg}. In LRSM, as the electroweak symmetry is broken dynamically, an accessible amount of LFV is predicted in a large region of the parameter space of the model \cite{Das:2012ii}. There are several lepton flavor violating decays, but the most relevant ones are the rare muon leptonic decays $\mu\rightarrow3e$ and $\mu\rightarrow e\gamma$. The relevant branching ratios (BR) for the processes are given as \cite{Borgohain:2017inp},
	\begin{equation}
		{BR}_{\mu\rightarrow3e} \equiv \frac{\Gamma(\mu^{+}\rightarrow e^{+}e^{-}e^{+})}{\Gamma_{\nu}}.
	\end{equation}
	\begin{equation}
		{BR}_{\mu\rightarrow e\gamma} \equiv \frac{\Gamma(\mu^{+}\rightarrow e^{+}\gamma )}{\Gamma_{\mu}}.
	\end{equation}
	The best upper limit for these branching ratios is set by the MEG collaboration \cite{MEG:2016leq} and SINDRUM experiment \cite{SINDRUM:1985vbg}, which provide the upper limit for $BR(\mu\rightarrow3e)$ $<$ $1.0 \times 10^{-12}$ and for $BR(\mu\rightarrow e\gamma)$ $<$ $4.2 \times 10^{-13}$. The branching ratios for these decays are given as follows:
	\begin{itemize}
		\item For the decay $BR(\mu\rightarrow3e)$, the BR is given as,
		\begin{equation}
			\label{x48}
			{BR}_{\mu\rightarrow3e} = \frac{1}{2}|h_{\mu e}h_{ee}^*|^2 \Biggl(\frac{{m_{W_L}}^4}{M_{{\Delta_L}^{++}}^{4}} + \frac{{m_{W_R}}^4}{M_{{\Delta_R}^{++}}^{4}}\Biggl)
		\end{equation}
		where, $h_{ij}$ describes the respective lepton-scalar couplings given by,
		\begin{equation}
			\label{x49}
			h_{ij} = \sum_{n=1}^{3} V_{in}V_{jn}\Biggl(\frac{M_n}{M_{W_{R}}}\Biggl)
		\end{equation}
		where, $i,j = e,\mu,\tau$.
		\item For the decay $BR(\mu\rightarrow e\gamma)$, the BR is given as,
		\begin{equation}
			\label{x50}
			{BR}_{\mu\rightarrow e\gamma} = 1.5 \times 10^{-7}|g_{lfv}|^2 \Biggl(\frac{1TeV}{M_{W_{R}}}\Biggl)^4
		\end{equation}
		where, \begin{equation}
			\label{x51}
			g_{lfv} = \sum_{n=1}^{3} V_{\mu n}{V_{en}}^{*}\Biggl(\frac{M_n}{M_{W_{R}}}\Biggl)^2 = \frac{[M_R {M_R}^{*}]_{\mu e}}{M_{W_{R}}}
		\end{equation}
	\end{itemize}
	The sum being over heavy neutrino. $V$ is the right-handed neutrino mixing matrix and $M_{{\Delta}_{L,R}}^{++}$ are the mass of the doubly charged boson.
	
	\begin{center}
		\section{\label{lrsm14}\textbf{Numerical Analysis and Results}}
	\end{center}

\subsection{\underline{Resonant leptogenesis with different neutrino mass textures in modular LRSM}}
\begin{figure}[H]
		\centering
		\includegraphics[scale=0.5]{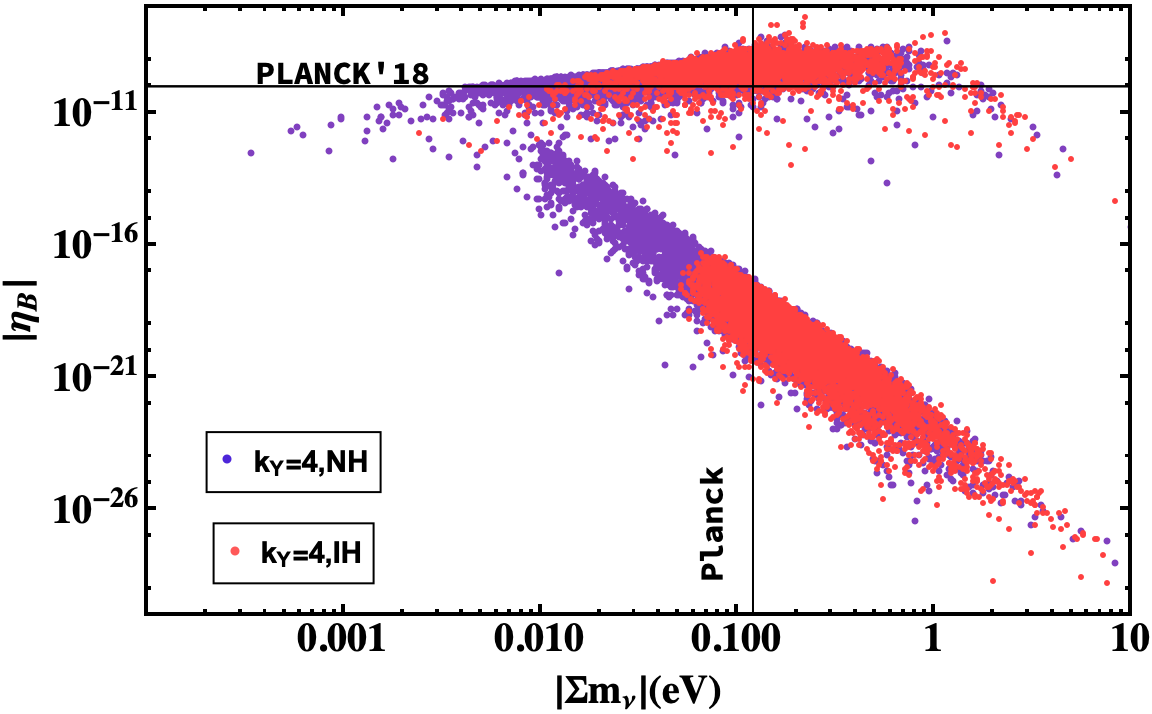}
		\caption{\label{f5}Variation of sum of neutrino masses with baryon asymmetry parameter for both NH and IH for weight 4.}
	\end{figure}

\begin{figure}[H]
		\centering
		\includegraphics[scale=0.5]{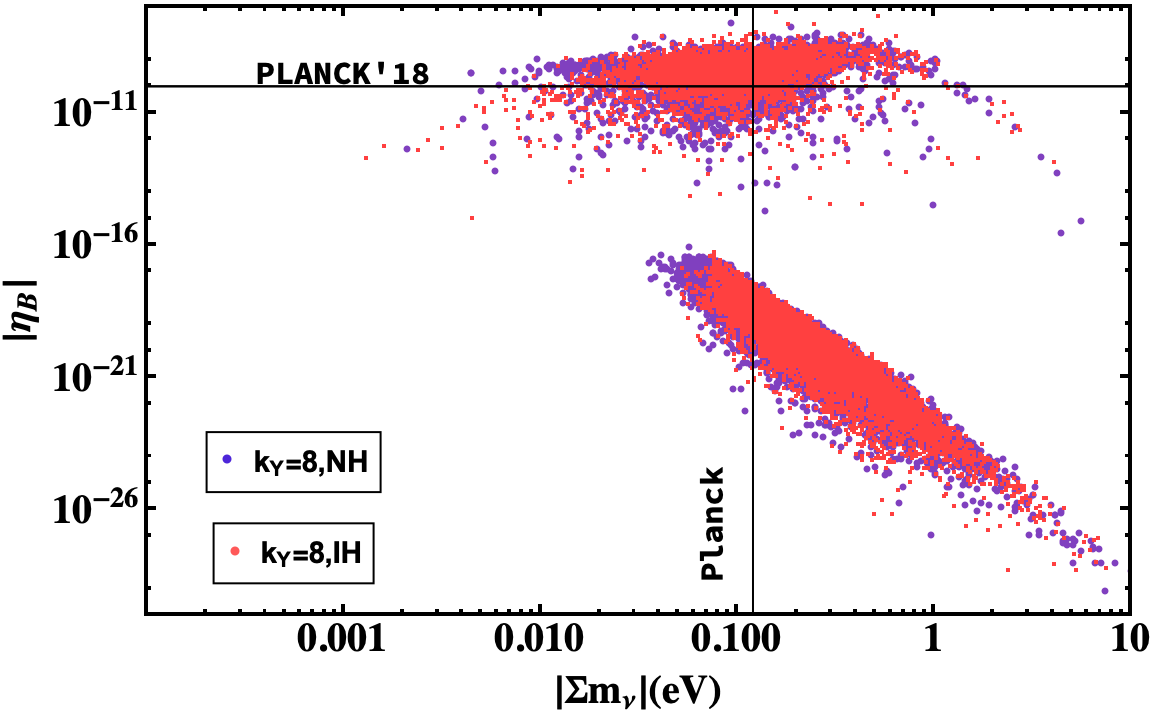}
		\caption{\label{f6}Variation of sum of neutrino masses with baryon asymmetry parameter for both NH and IH for weight 8.}
	\end{figure}

\begin{figure}[H]
	\centering
	\includegraphics[scale=0.38]{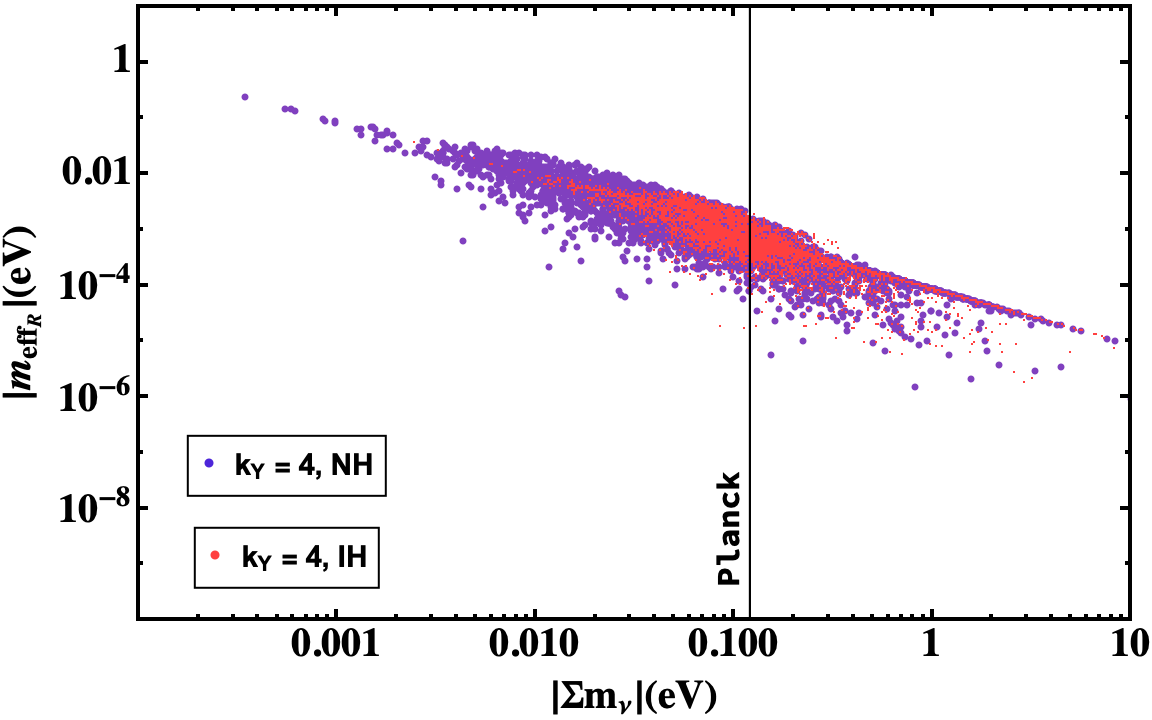}
	\includegraphics[scale=0.38]{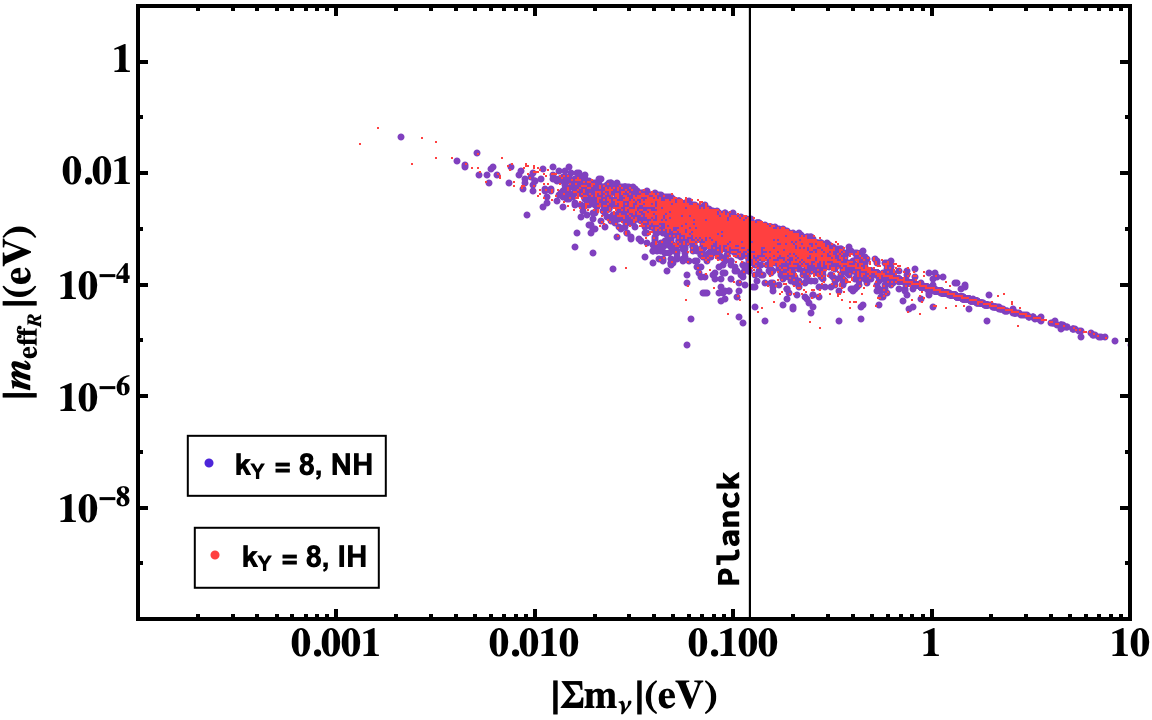}
	\caption{\label{f7}Variation of sum of neutrino masses with effective mass for heavy right-handed contribution for weight 4(left) and weight 8(right).}
\end{figure}

\begin{figure}[H]
	\centering
	\includegraphics[scale=0.38]{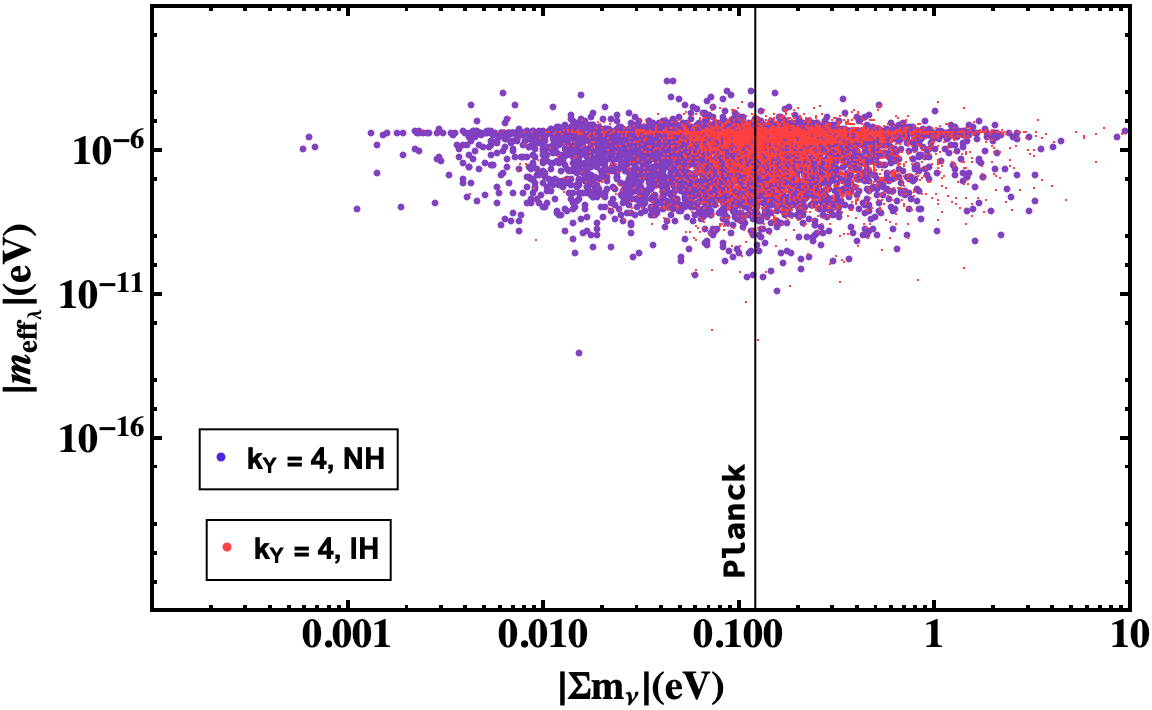}
	\includegraphics[scale=0.38]{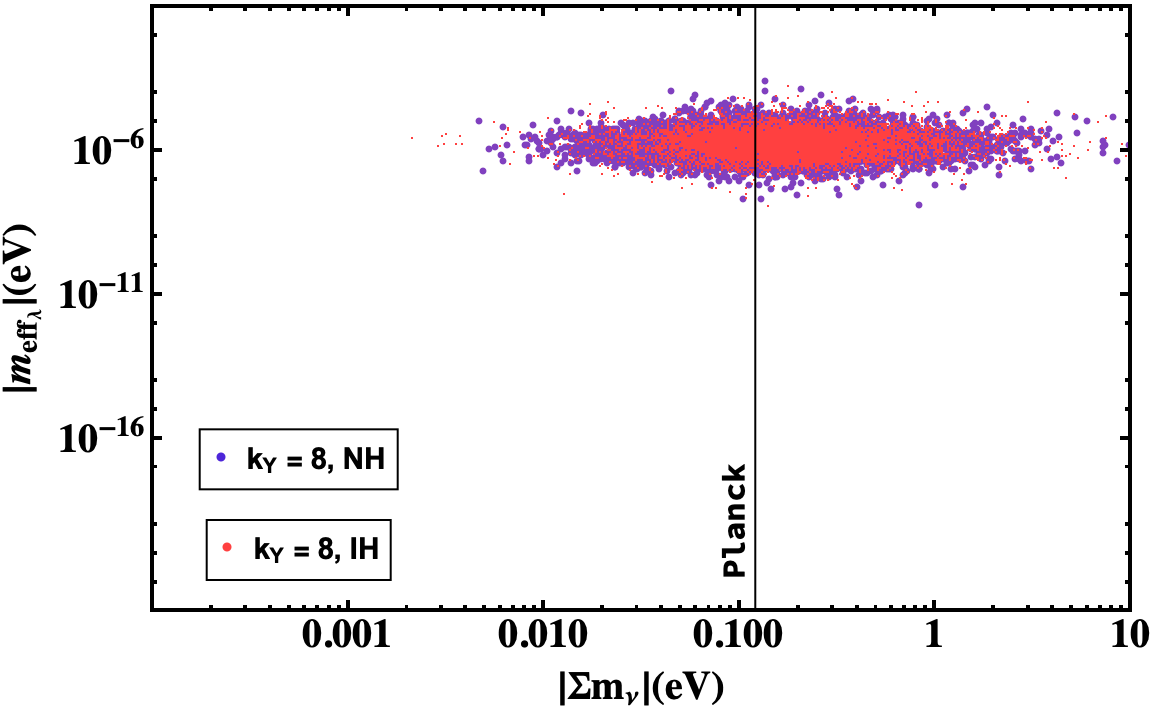}
	\caption{\label{f8}Variation of sum of neutrino masses with effective mass for $\lambda$ contribution for weight 4(left) and weight 8(right).}
\end{figure}

\begin{figure}[H]
	\centering
	\includegraphics[scale=0.38]{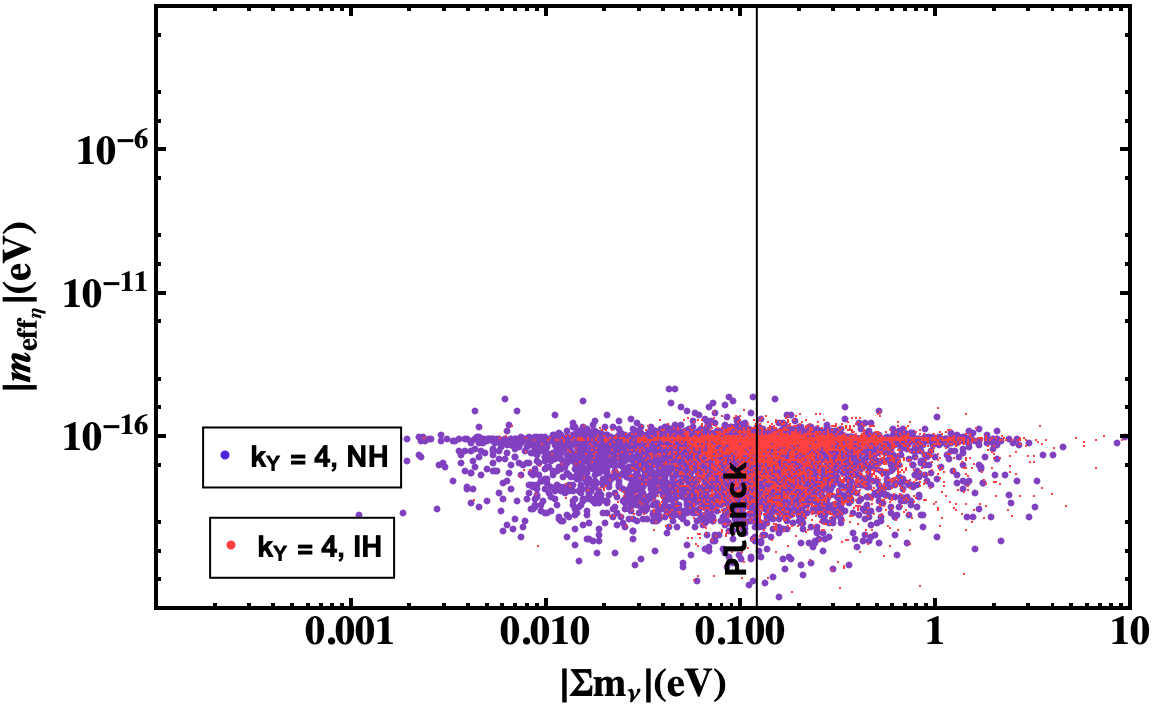}
	\includegraphics[scale=0.38]{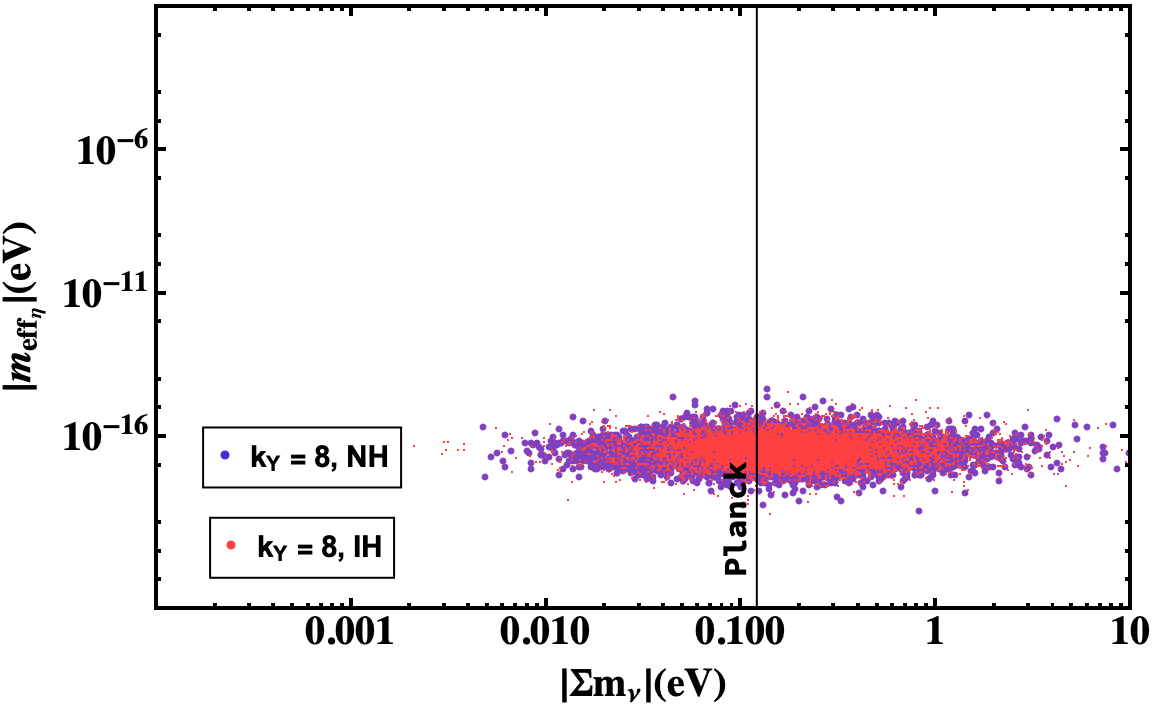}
	\caption{\label{f9}Variation of sum of neutrino masses with effective mass for $\eta$ contribution for weight 4(left) and weight 8(right).}
\end{figure}

	The leptonic CP asymmetry $\epsilon_{1}$ and $\epsilon_{2}$ are determined by using \eqref{x46}, where $Y_v = \frac{M_D}{v}$, $M_D$ being the Dirac mass and $v$ is the VEV of the Higgs bidoublet and is in $GeV$. The decay rates in \eqref{x47} can be calculated using the relation \eqref{x46}.\\
	The CP violating asymmetries $\epsilon_{1}$ and $\epsilon_{2}$ can give rise to net lepton number asymmetry, provided the expansion rate of the universe is larger than $\Gamma_{1}$ and $\Gamma_{2}$.The net baryon asymmetry is then calculated using the following relation \cite{Buchmuller:2004tu},
	\begin{equation}
		\label{x52}
		\eta_B \approx -0.96 \times 10^{-2}\sum_{i}(k_{i}\epsilon_{i})
	\end{equation}
	$k_i$ being the efficiency factors measuring the washout effects. Some parameters are needed to defined as,
	\begin{equation}
		\label{x53}
		K_i \equiv \frac{\Gamma_{i}}{H}
	\end{equation}
	Equation \eqref{x53} is defined at temperature $T=M_{i}$. The Hubble's constant is given by, $H\equiv \frac{1.66\sqrt{g_{*}}T^{2}}{M_{Planck}}$, where, $g_{*}=107$ and $M_{Planck} = 1.2 \times 10^{19}$ GeV is the Planck mass. Decay width is estimated using \eqref{x46}. The efficiency factors $k_i$ can be calculated using the formula \cite{Blanchet:2008pw},
	\begin{equation}
		\label{x54}
		k_1 \equiv k_2 \equiv \frac{1}{2}(\Sigma_{i} K_{i})^{-1.2}
	\end{equation}
	The results obtained for leptogenesis for both the cases $k=4$ and $k=8$ are shown in figures \ref{f5} and \ref{f6}. In figures \ref{f5} to \ref{f17}, NH refers to normal hierarchy and IH to inverted hierarchy.

	\subsection{\underline{$0\nu\beta\beta$ with different neutrino mass textures in modular LRSM}}
	As already stated, $0\nu\beta\beta$ in LRSM has eight contributions in total, however in the present work, we have considered the the new physics contribution corresponding to the exchange of heavy right-handed neutrino, and the momentum dependent $\lambda$ and $\eta$ contributions. \\
	
	The effective mass corresponding to heavy right-handed neutrino contribution is given by,
	\begin{equation}
		m_R^{eff} = p^2\Bigg({\frac{M_{W_L}}{M_{W_R}}}\Bigg)^{4}{\frac{U_{Rei}^{*}{^2}}{M_i}}
	\end{equation}
	where, $p^2$ is the typical momentum exchange of the process. As it is known that TeV scale LRSM plays a very important role in the process of neutrinoless double beta decay (0$\nu\beta\beta$), we have considered the values as $M_{W_R} = 10TeV$ , $M_{W_L} = 80GeV$ , $M_{\Delta_R} \approx 3TeV$  and after calculation, the value for heavy right-handed neutrino is found to be in the scale of $TeV$. The allowed value of p is in the range $(100-200)MeV$ and so we consider, $p \approx 180 MeV$.
	Thus, we get,
	\begin{equation}
		p^2\Bigg(\frac{M_{W_L}}{M_{W_R}}\Bigg)^{4} = 10^{10} eV^{2}
	\end{equation}
	where, $U_{Rei}$ refers to the first row elements of the diagonalizing matrix and $M_i$ are its eigenvalues. The momentum dependent contributions are described by the respective dimensionless parameters
	\begin{itemize}
		\item For $\lambda$ contribution, the dimensionless parameter $\eta_\lambda$ is given by,
		\begin{equation}
			\label{x55}
			|{\eta_\lambda}| = \Biggl(\frac{M_{W_L}}{M_{W_R}}\Biggl)^{2}|\Sigma_{i}U_{ei}T_{ei}^*|
		\end{equation}
		\item For $\eta$ contribution, the dimensionless parameter describing $0\nu\beta\beta$ is given by,
		\begin{equation}
			\label{x56}
			|\eta_{\eta}| = \tan \xi |\Sigma_{i}U_{ei}T_{ei}^*|
		\end{equation}
	\end{itemize}
	In the above equations, $U_{ei}$ represents the first row of the neutrino mixing matrix. $|\Sigma_{i}U_{ei}T_{ei}^*|$ can be simplified to the form $-[M_{D}M_{RR}^{-1}]_{ee}$ as described in \cite{Barry:2013xxa}. T is represented by the equation \eqref{x17} and,
	\begin{equation}
		\label{x57}
		tan 2\xi = -\frac{2kk'}{v_{R}^2 - v_{L}^2}
	\end{equation}
	The respective effective masses are expressed as,
	\begin{equation}
		\label{x58}
		\centering
		|m_{eff_{\lambda}}|=|\eta_\lambda||m_{e}|;|m_{eff_{\eta}}|=|\eta_\eta||m_{e}|
	\end{equation}
	where, $m_{e}$ is the mass of the electron.
	\subsection{\underline{Lepton Flavor Violation with different neutrino mass textures in modular LRSM}}
\begin{figure}[H]
		\centering
		\includegraphics[scale=0.43]{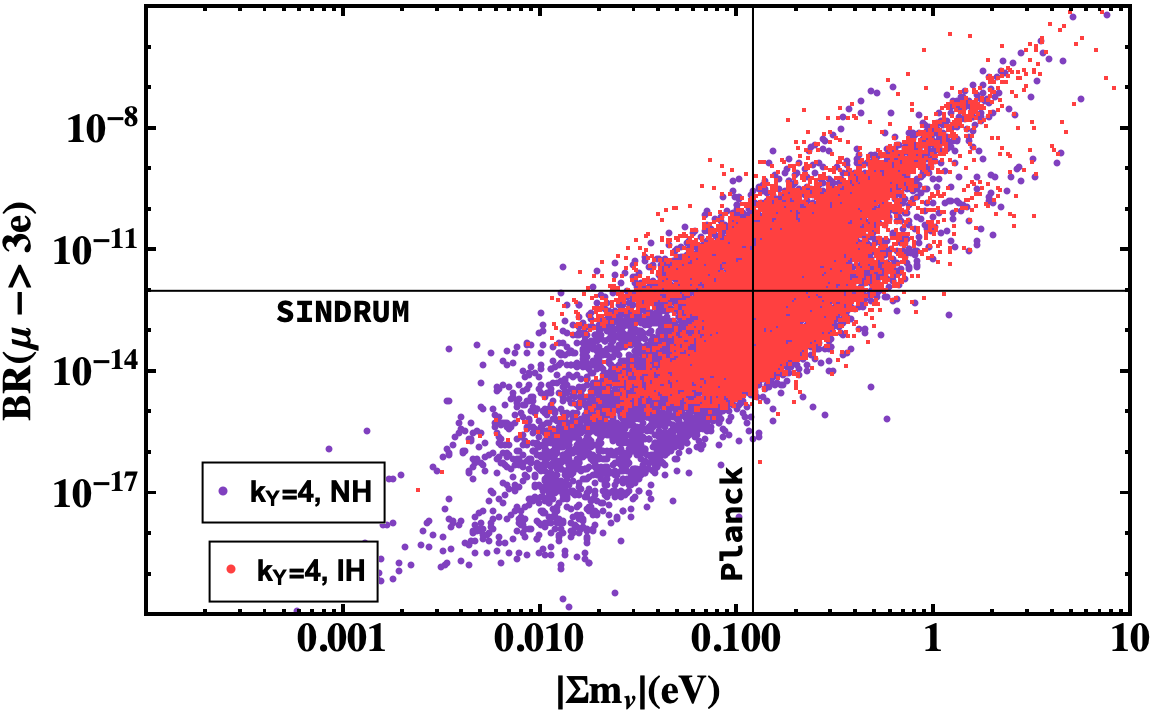}
		\includegraphics[scale=0.43]{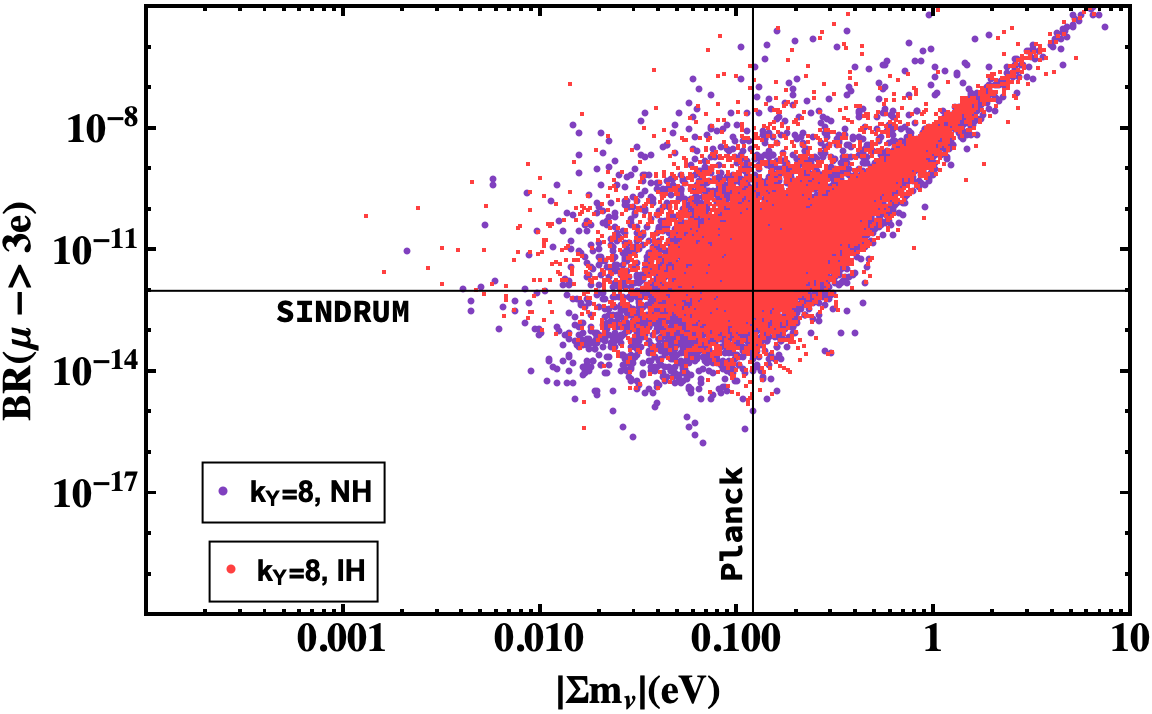}
		\caption{\label{f11}Variation of sum of neutrino masses with branching ratio for $\mu \rightarrow 3e$ for weights 4(left) and 8(right).}
	\end{figure}

\begin{figure}[H]
	\centering
	\includegraphics[scale=0.43]{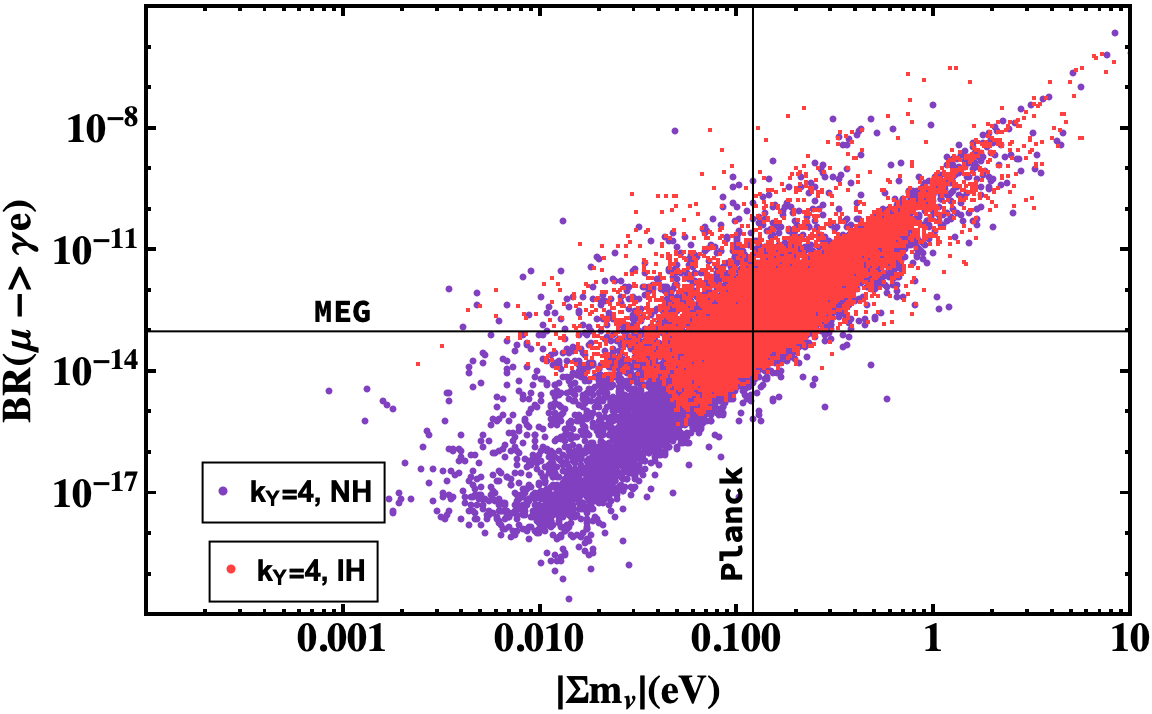}
	\includegraphics[scale=0.43]{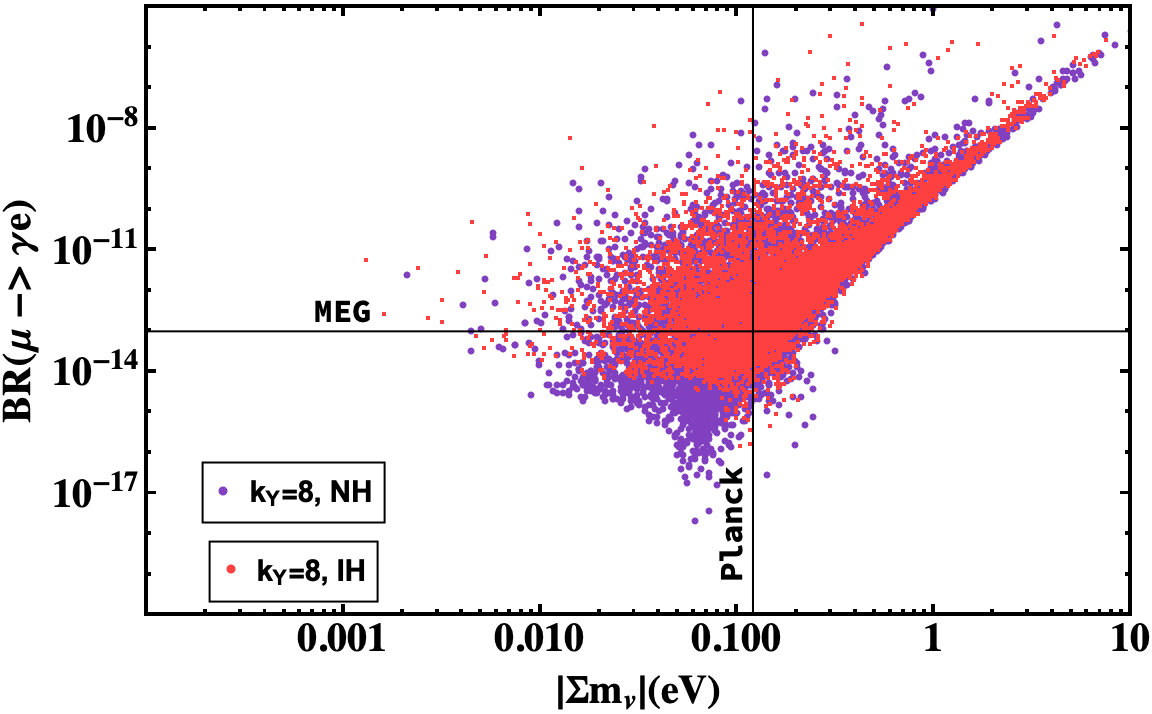}
	\caption{\label{f12}Variation of sum of neutrino masses with branching ratio for $\mu \rightarrow \gamma e$ for weights 4(left) and 8(right).}
\end{figure}

	As already stated that the most relevant lepton flavor violating decays are the rare muon leptonic decays $\mu\rightarrow3e$ and $\mu\rightarrow e\gamma$. The branching ratio for the same has been calculated using the relation \eqref{x48} to \eqref{x51}. The result for the same are shown in figures \ref{f11} and \ref{f12} for both the decays.
\subsection{\underline{Resonant leptogenesis, $0\nu\beta\beta$ and LFV  in modular LRSM}}
\begin{figure}[H]
	\centering
	\includegraphics[scale=0.43]{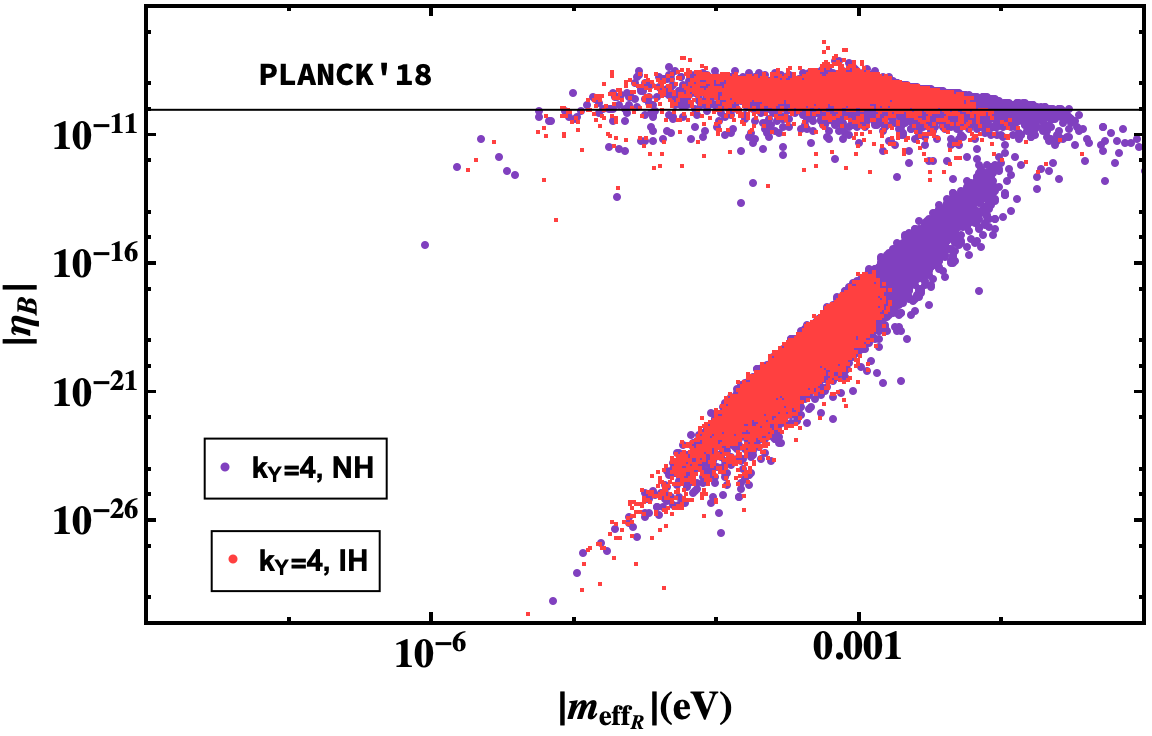}
	\includegraphics[scale=0.43]{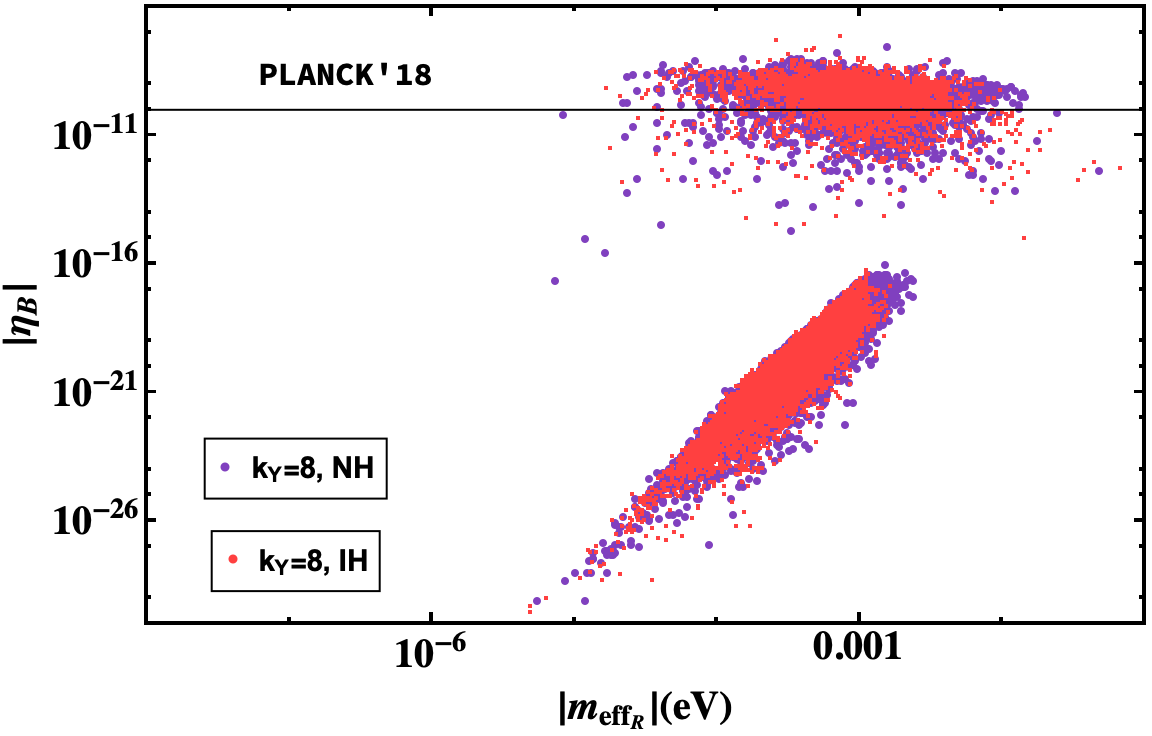}
	\caption{\label{f13}Variation of effective mass for heavy right-handed neutrino contribution with baryon asymmetry parameter for weights 4(left) and 8(right).}
\end{figure}

\begin{figure}[H]
	\centering
	\includegraphics[scale=0.43]{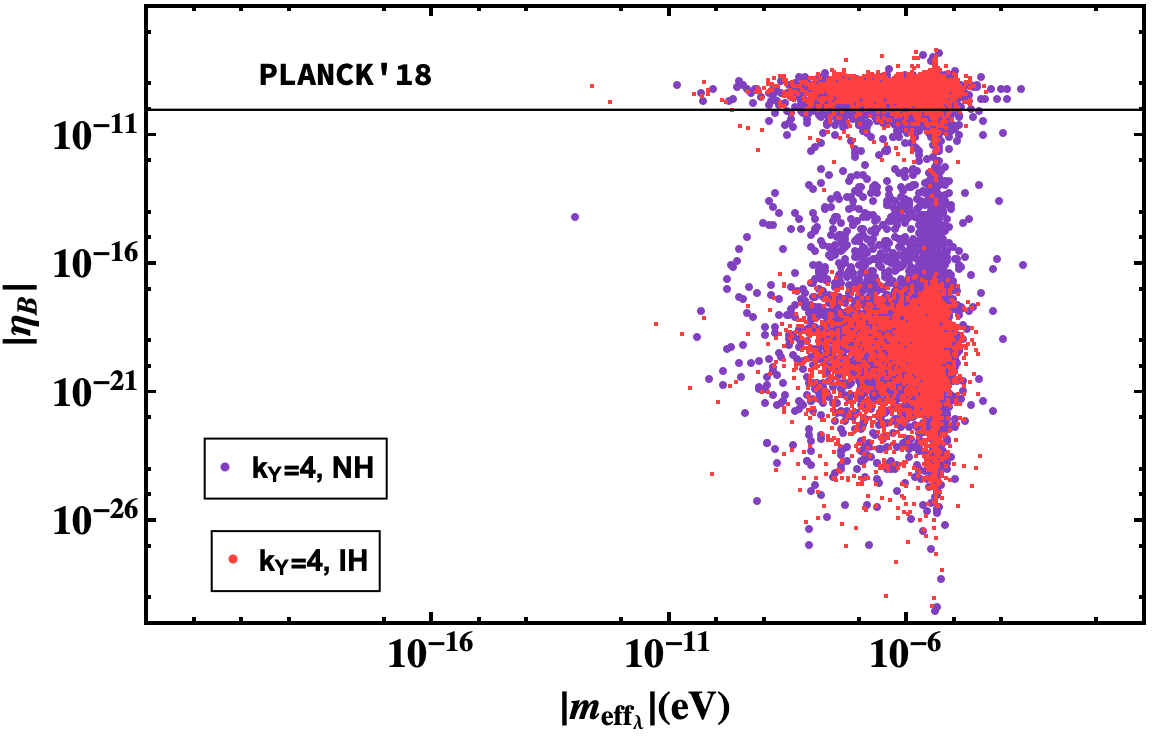}
	\includegraphics[scale=0.43]{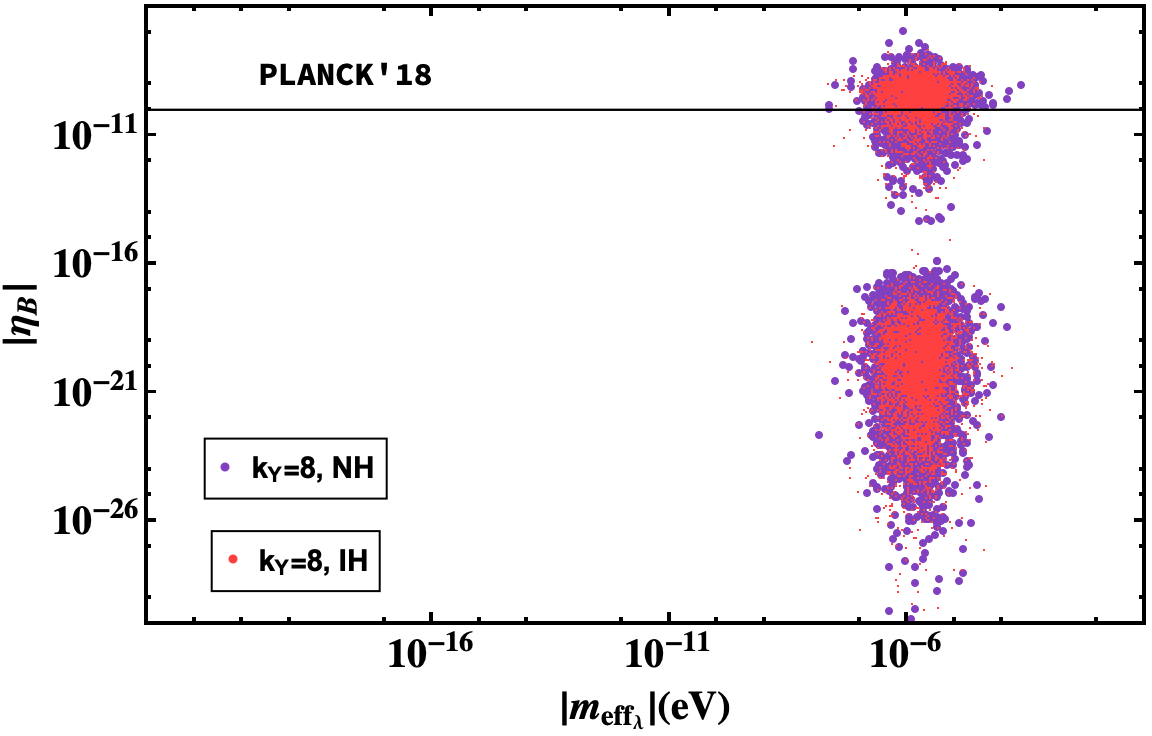}
	\caption{\label{f14}Variation of effective mass for $\lambda$ contribution with baryon asymmetry parameter for weights 4(left) and 8(right).}
\end{figure}

\begin{figure}[H]
	\centering
	\includegraphics[scale=0.43]{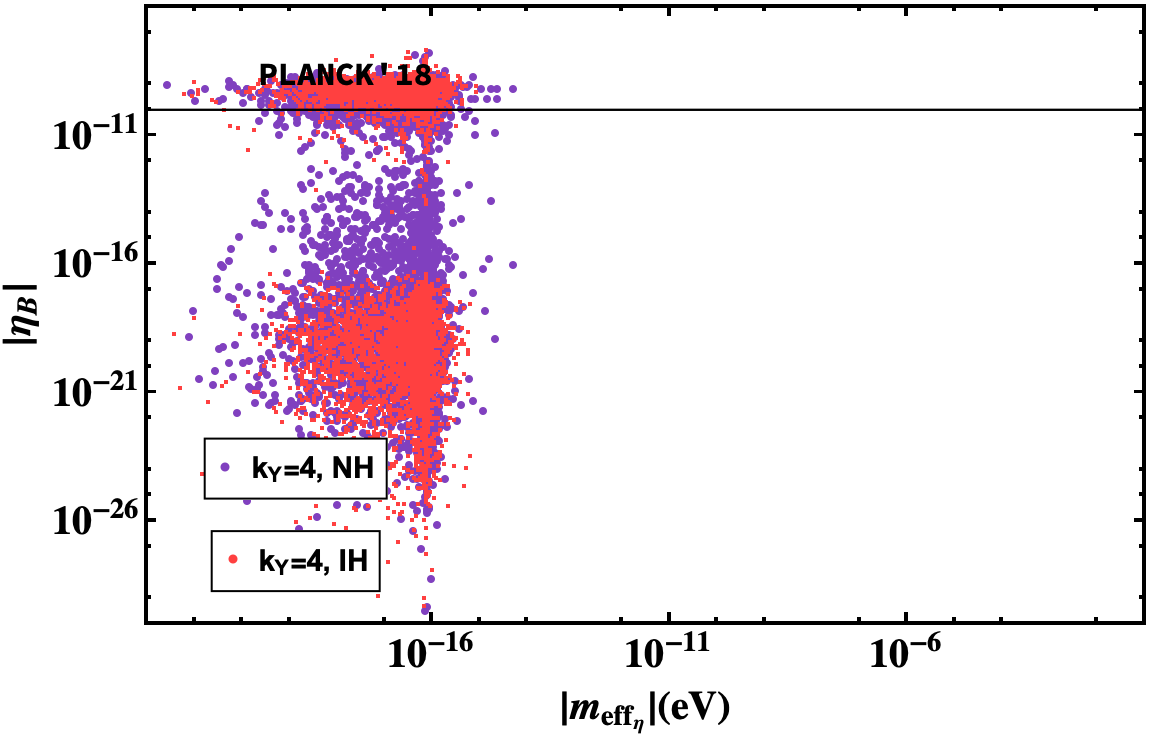}
	\includegraphics[scale=0.43]{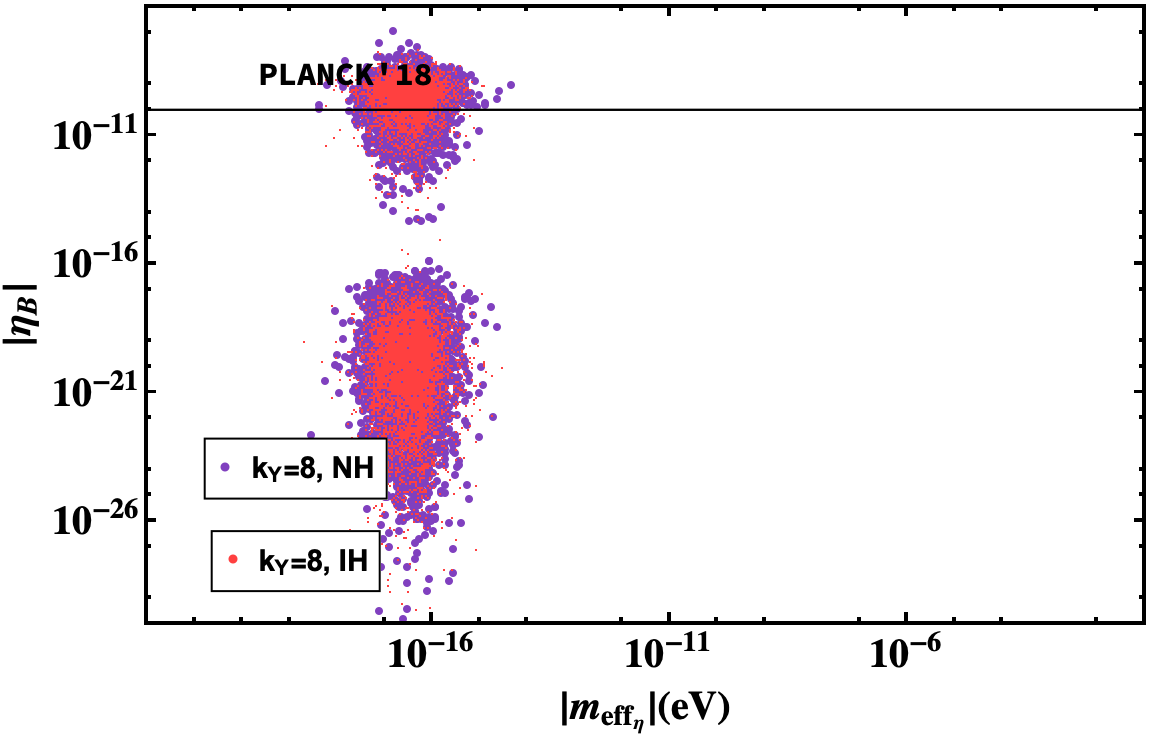}
	\caption{\label{f15}Variation of effective mass for $\eta$ contribution with baryon asymmetry parameter for weights 4(left) and 8(right).}
\end{figure}

\begin{figure}[H]
	\centering
	\includegraphics[scale=0.43]{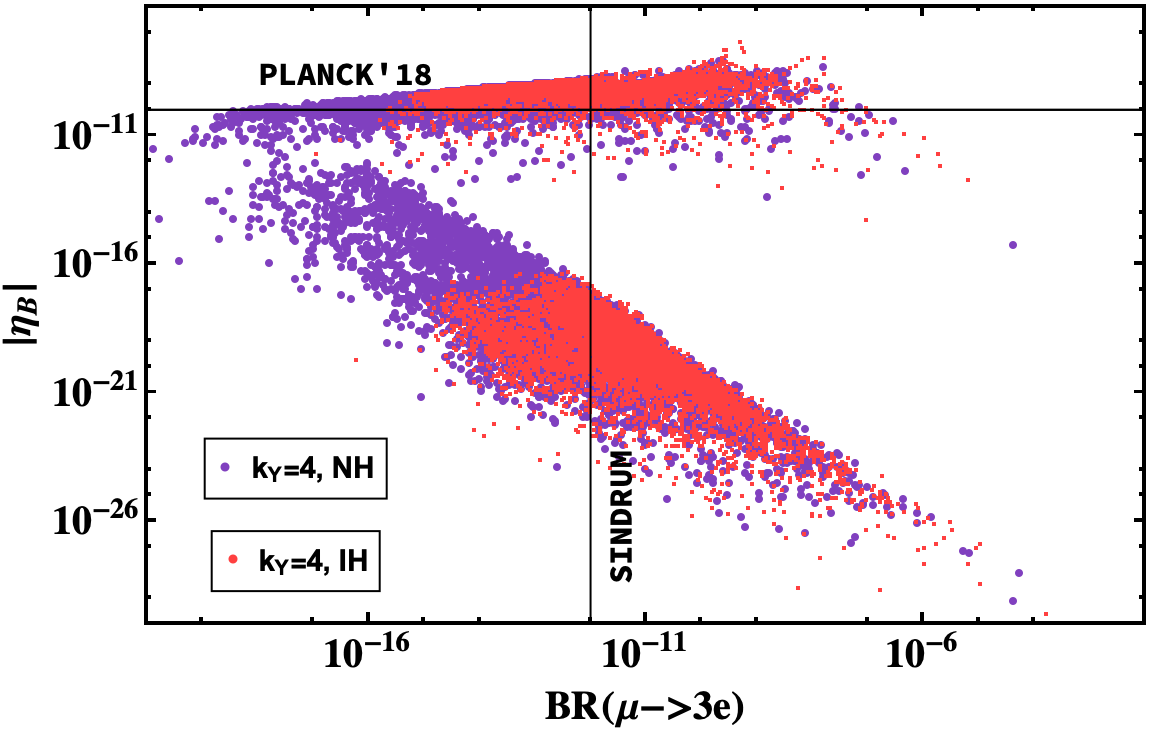}
	\includegraphics[scale=0.43]{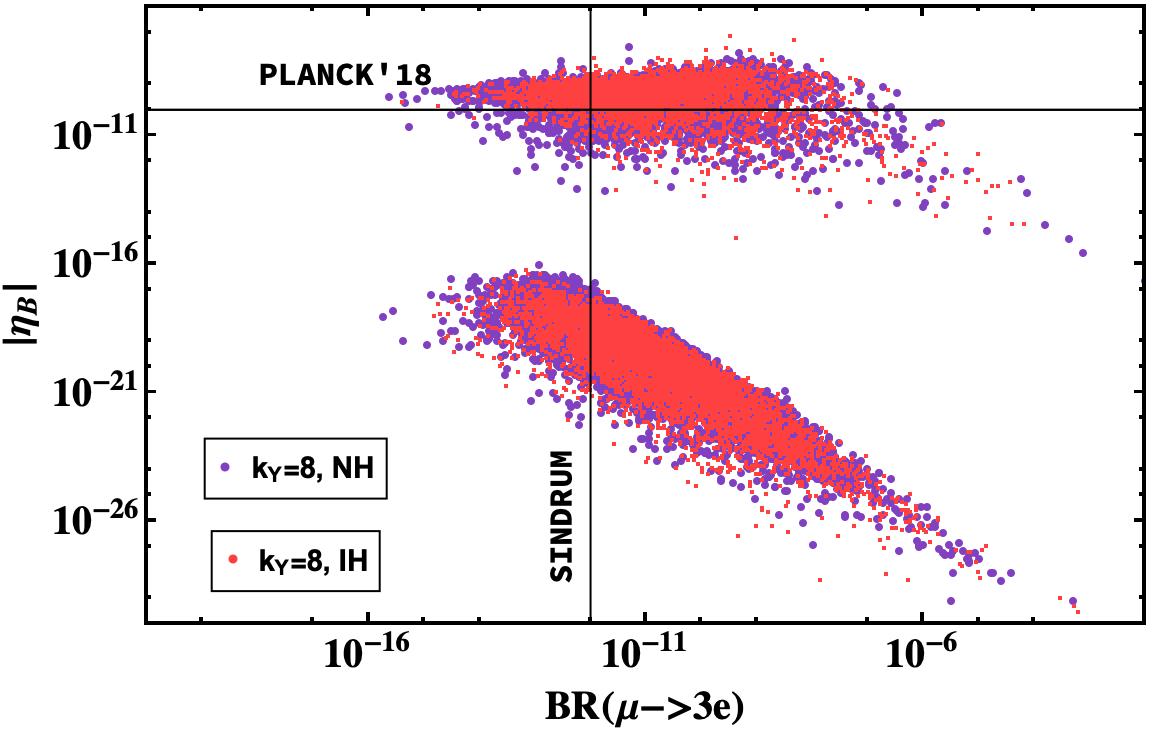}
	\caption{\label{f16}Variation of branching ratio for $\mu \rightarrow 3e$ with baryon asymmetry parameter for weights 4(left) and 8(right).}
\end{figure}

\begin{figure}[H]
	\centering
	\includegraphics[scale=0.43]{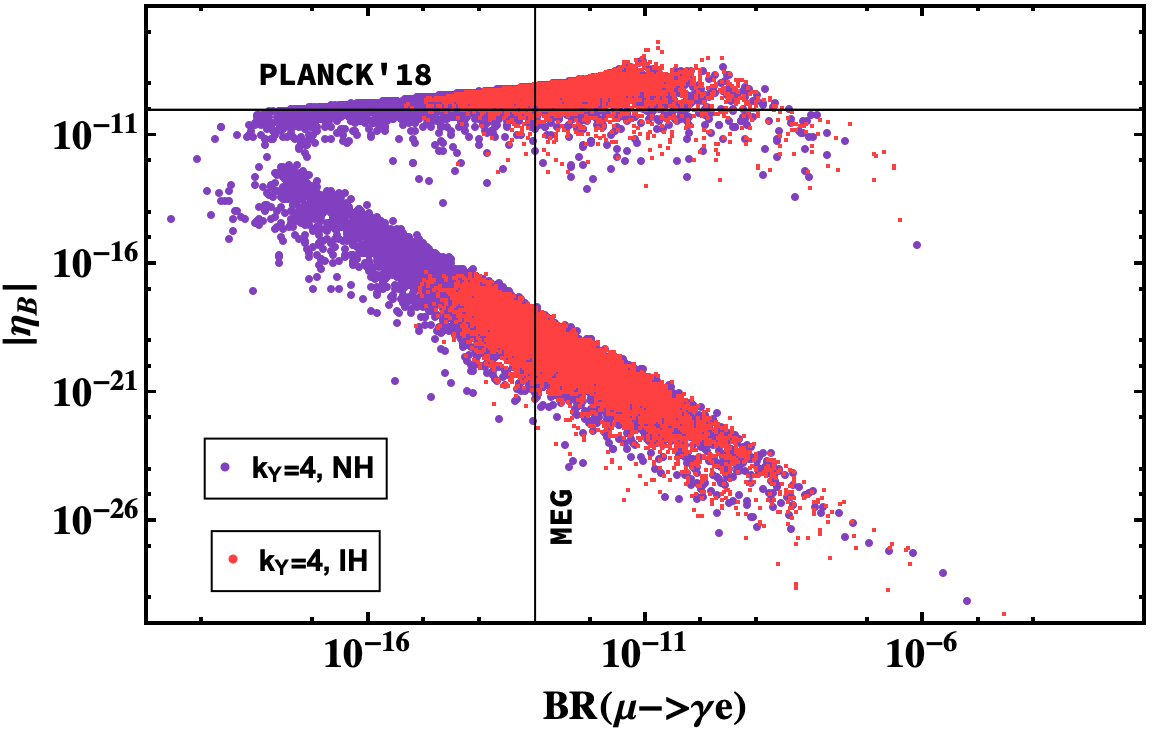}
	\includegraphics[scale=0.43]{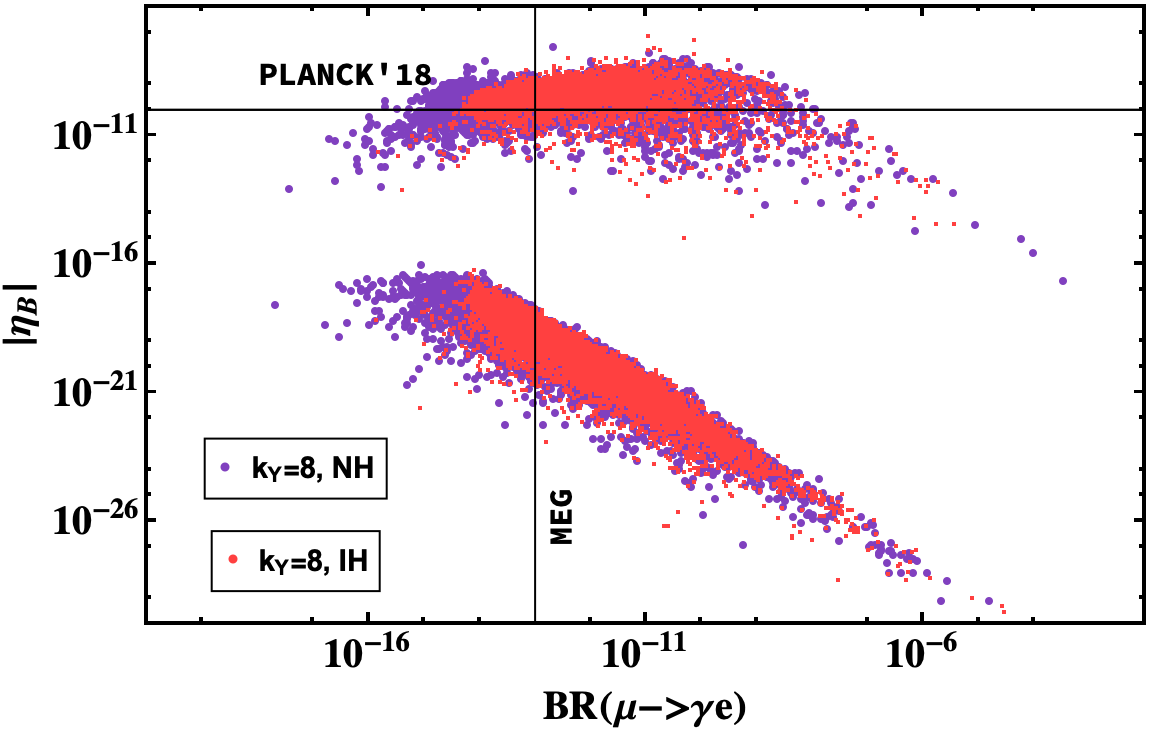}
	\caption{\label{f17}Variation of branching ratio for $\mu \rightarrow \gamma e$ with baryon asymmetry parameter for weights 4(left) and 8(right).}
\end{figure}

Figures relating different phenomenological parameters with each other have also been studied. The plots relating neutrinoless double beta decay with resonant leptogenesis have been shown in figures \ref{f13} to \ref{f16} for weights $k_{Y}=4$ and $k_{Y}=8$.
\newpage
The plots showing the variations of lepton flavor violation with resonant leptogenesis are shown in figures \ref{f16} to \ref{f17}.

	\begin{center}
		\section{\label{lrsm15}Discussion and Conclusion}
	\end{center}
	The realization of LRSM with $A_4$ modular symmetry provides the advantage of not requiring the use of any extra particles (flavons) for acquiring the desired results. We are emphasizing on different textures of the neutrino mass matrices arising as a result of implementing different weights to the repective modular forms. For phenomenological studies, we have taken into consideration resonant leptogenesis, $0\nu\beta\beta$ and lepton flavor violation. Given below is a thorough discussion of our current work and the important results obtained for the same.
	\begin{itemize}
		\item Left-Right symmetric model in the present work has been realized with the help of modular group of level 3 ($\Gamma(3)$). Corresponding to the level $N$ of the modular group, the number of modular forms are determined. If $2 \leq N \leq 5$, then the corresponding modular group is isomorphic to a particular non-abelian discrete symmetry group as shown in table \ref{table:1}. Hence, in the current work we realize LRSM with $A_4$ non-abelian discrete symmetry group for it is isomorphic to $\Gamma(3)$ modular group.
		\item While using modular symmetry, the Yukawa couplings within a particular model are expressed as modular forms, say $Y_{i}'s$, where their number is determined by the level and weight of the modular group used. In the present work, $\Gamma(3)$ modular group of weights 4, 8 and 10 which corresponds to five, nine  and eleven modular forms respectively as depicted in expressions \eqref{x20} to \eqref{x35}. where the modular forms are expressed in terms of  $(Y_1,Y_2,Y_3)$, and they are functions of $q$ which is expressed as $q=\exp (2\pi i \tau)$, $\tau$ being the complex modulus in the upper half of the complex plane.
		\item After construction of the Yukawa Lagrangian for modular LRSM, taking into consideration all the weights as discussed in sections \ref{lrsm112} and \ref{lrsm114a}  we determine the Dirac and Majorana mass matrices in terms of the modular forms. This helps us in determining the values for unknown $(Y_1,Y_2,Y_3)$, and hence determine the value of $\tau$.  The modular parameters are found to satisfy the fundamental domain , that is, $|Re(\tau)| \leq 0.5$, $Im(\tau) > 0$ and $|\tau| >1$.  For calculation, we have used $3\sigma$ values of the neutrino oscillation parameters as given in \cite{Esteban:2020cvm}.
		\item For LRSM, the resulting light neutrino mass is given as a summation of type-I and type-II seesaw masses and as such, the origin of different textures in the neutrino mass matrix  depends upon the charge assignments and in the current case, on the modular weights of the particle content. We have considered weights $k_{Y}=4$, $k_{Y}=8$ and $k_{Y}=10$ for realizing the neutrino mass matrix. As described in expressions \ref{x19} to \ref{x35}, modular forms of different weights have different irreducible representations and all of them can be expressed in terms of  $(Y_1,Y_2,Y_3)$ and as such our resulting light neutrino mass matrix is expressed in terms of the same.
		\item The resulting light neutrino mass matrix in our case is symmetric in nature for all the cases. Considering weight $k_{Y}=4$, it has been seen that we arrive at Class $B_2$ of 2-0 texture, when $k=6$, there is no possible outcome for 2-0 texture in the neutrino mass matrix. When $k_{Y}=8$, considering the modular forms as singlets, we arrive at the mass matrix given by \eqref{x43}. Taking different combinations of modular weights for the modular form in the same Lagrangian may help us arrive at some other textures in the neutrino mass matrix, but this might result in the addition of some free parameters within the Lagrangian and also some additional terms, for which we have considered only one weight for the modular form at one particular case, in order to keep the analysis minimal.
		\item Coming to the study of phenomenology associated with  TeV scale LRSM, we take into account the study of resonant leptogenesis (RL), neutrinoless double beta decay $(0\nu\beta\beta)$ and lepton flavor violation (LFV). Talking about RL, favourable results are seen for different classes of neutrino mass textures in case of both normal and inverted hierarchy as depicted in figures \ref{f5} and \ref{f6}.
		\item  For $0\nu\beta\beta$, we have considered heavy right-handed neutrino and the momentum dependent $\lambda$ and $\eta$ contributions, the results for which are found to satisfy the experimental bounds for both normal and inverted hierarchy.
		\item Coming to LFV, considering the decays $\mu \rightarrow 3e$, $\mu \rightarrow \gamma e$ for both the cases of weights 4 and 8,  it has been observed that the results satisfy the experimental bounds for both SINDRUM and MEG experiments respectively for both normal and inverted hierarchy. 
		\item  The main purpose of our work is to search for a common parameter space where we can correlate RL, $0\nu\beta\beta$ and LFV at a TeV scale LRSM which obeys $A_4$ modular symmetry. From our analysis, we see that one can account for successful leptogenesis, neutrinoless double beta decay and also LFV in such a scenario for the different textures obtained for the neutrino mass matrix. If we talk about a common parameter space for the three phenomenology considered in the present work, the results for all of the same are found to be satisfied when the sum of the neutrino masses lies within a range of $10^{-2}$ to approximately $0.5$ eV.
	\end{itemize}
	LRSM which is a simple extension of the ideal Standard Model of particle physics stands a suitable platform for simultaneous study of different phenomenology together, as is evident from the present work. Several previous works have been done in the model taking flavor symmetry into consideration\cite{Boruah:2022bvf,Rodejohann:2015hka,Sahu:2020tqe,Boruah:2021ktk}. However, modular symmetry has proven to be favourable because of the least number of free parameters and it has also been found to be fruitful in the phenomenological specification of allowed and disallowed cases of texture zeros in neutrino mass matrix. Some previous work as dicussed in \cite{Borgohain:2017akh,Borgohain:2017inp} have also taken into account the study of $0\nu\beta\beta$ with changing values of $M_{W_R}$, and in our work \cite{Kakoti:2023xkn}, we have taken different values of $M_{W_R}$ and also different strengths of type-II seesaw mass into consideration. In the current TeV scale realization of LRSM, we have used the value $M_{W_R}=10TeV$, and the results are found to be as shown and discussed above. So, conclusively we can use modular symmetry for the realization of LRSM with no extra particles and hence study different phenomenology within the usual particle content of the model which helps in keeping the model minimal. Interestingly, our work prefers both normal and inverted ordering of neutrino mass when different phenomenological study are considered with some distinctions in the observed parameter space within the framework of modular LRSM. The model can also be tested at experiments owing to its study in TeV scale LRSM.

	\begin{center}
		\section*{Appendix A : Properties of $A_4$ discrete symmetry group.}
	\end{center}
	
	$A_4$ is a non-abelian discrete symmetry group which represents even permuatations of four objects. It has four irreducible representations, three out of which are singlets $(1,1',1'')$ and one triplet $3$ ($3_A$ represents the anti-symmetric part and $3_S$ the symmetric part). Products of the singlets and triplets are given by,
	\begin{center}
		\begin{equation*}
			1 \otimes 1 = 1
		\end{equation*}
	\end{center}
	\begin{center}
		\begin{equation*}
			1' \otimes 1' = 1''
		\end{equation*}
	\end{center}
	\begin{center}
		\begin{equation*}
			1' \otimes 1'' = 1
		\end{equation*}
	\end{center}
	\begin{center}
		\begin{equation*}
			1'' \otimes 1'' = 1'
		\end{equation*}
	\end{center}
	\begin{center}
		\begin{equation*}
			3 \otimes 3 = 1 \oplus 1' \oplus 1'' \oplus 3_A \oplus 3_S
		\end{equation*}
	\end{center}
	If we have two triplets under $A_4$ say, $(a_1,a_2,a_3)$ and $(b_1,b_2,b_3)$ , then their multiplication rules are given by,
	\begin{center}
		\begin{equation*}
			1 \approx a_1b_1 + a_2b_3 + a_3b_2
		\end{equation*}
		\begin{equation*}
			1' \approx a_3b_3 + a_1b_2 + a_2b_1
		\end{equation*}
		\begin{equation*}
			1'' \approx a_2b_2 + a_3b_1 + a_1b_3
		\end{equation*}
		\begin{equation*}
			3_S \approx \begin{pmatrix}
				2a_1b_1-a_2b_3-a_3b_2 \\
				2a_3b_3-a_1b_2-a_2b_1 \\
				2a_2b_2-a_1b_3-a_3b_1
			\end{pmatrix}
		\end{equation*}
		\begin{equation*}
			3_A \approx \begin{pmatrix}
				a_2b_3-a_3b_2 \\
				a_1b_2-a_2b_1 \\
				a_3b_1-a_1b_3
			\end{pmatrix}
		\end{equation*}
	\end{center}

	\section*{Acknowledgements}
	
	Ankita Kakoti acknowledges Department of Science and Technology (DST), India (grant DST/INSPIRE Fellowship/2019/IF190900) for the financial assistantship.
	\vspace{1cm}
	\begin{center}
		\section*{References}
	\end{center}
	
	\bibliography{citenew}
	\bibliographystyle{utphys}

\end{document}